\documentclass[a4paper,fleqn,usenatbib]{mnras}

\usepackage[T1]{fontenc}
\usepackage{ae,aecompl}

\usepackage{graphicx}	% Including figure files
\usepackage{multirow}
\usepackage{pdflscape}
\usepackage{amssymb}
\usepackage{amsmath}	% Advanced maths commands
\usepackage{bm}
\usepackage{morefloats}
\usepackage{txfonts}
\usepackage{hyperref}
\usepackage{natbib}
\usepackage{threeparttable}
\newcommand{\hvc}{HVC}
\newcommand{\hvcs}{HVCs}

\newcommand{\ivcs}{IVCs}
\newcommand{\ism}{ISM}
\newcommand{\him}{HIM}
\newcommand{\cgm}{CGM}
\newcommand{\khi}{KHI}
\newcommand{\rti}{RTI}
\newcommand{\rcloud}{R_{\rm cloud}}
\newcommand{\mcloud}{M_{\rm cloud}}
\newcommand{\tcloud}{T_{\rm cloud}}
\newcommand{\rhocloud}{\varrho_{\rm cloud}}

\newcommand{\tcentre}{T_{\rm centre}}
\newcommand{\tcgm}{T_{\rm CGM}}
\newcommand{\ncgm}{n_{\rm CGM}}
\newcommand{\rhocgm}{\varrho_{\rm CGM}}
\newcommand{\mmax}{M_{\rm BE}}
\newcommand{\mjeans}{M_{\rm J}}
\newcommand{\cs}{c_{\rm s}}
\newcommand{\lf}{\lambda_{\rm F}}
\newcommand{\lel}{\lambda_{\rm e}}
\newcommand{\LT}{L_{\rm T}}
\newcommand{\kms}{\,\textrm{km}\,\textrm{s}^{-1}}
\newcommand{\zsolar}{\,\textrm{Z}_\odot}
\newcommand{\msolar}{\,\textrm{M}_\odot}
\newcommand{\scm}{\,\textrm{cm}^{-2}}
\newcommand{\ccm}{\,\textrm{cm}^{-3}}

\title[Compact high-velocity clouds in the CGM]{Physical effects on compact high-velocity clouds in the circumgalactic medium}

\author[B. Sander \& G. Hensler]{
Bastian Sander,$^{1,2}$\thanks{E-mail: bastian.sander@iff.fraunhofer.de, gerhard.hensler@univie.ac.at}
Gerhard Hensler,$^{1}$\footnotemark[1]
\\
$^{1}$Department of Astrophysics, University of Vienna, T\"urkenschanzstra\ss e 17, A-1180 Vienna, Austria\\
$^{2}$Fraunhofer Institute for Factory Operation and Automation IFF, Sandtorstra\ss e 22, 39106 Magdeburg, Germany
}

\date{Accepted XXX. Received YYY; in original form ZZZ}

\pubyear{2020}

\begin{document}
\label{firstpage}
\pagerange{\pageref{firstpage}--\pageref{lastpage}}
\maketitle

\begin{abstract}
We numerically investigate the evolution of compact high-velocity clouds (C\hvcs) passing through a hot, tenuous gas representing the highly-ionized circumgalactic medium (\cgm) by applying the adaptive-mesh refinement code {\sc Flash}.

The model clouds start from both hydrostatic and thermal equilibrium and are in pressure balance with the \cgm. Here, we present $14$ models, divided into two mass categories and two metallicities each and different velocities. We allow for self-gravity and thermal conduction or not. All models experience mass diffusion, radiative cooling, and external heating leading to dissociation and ionization.

Our main findings are: 1) self-gravity stabilizes clouds against Rayleigh-Taylor instability which are disrupted within $10$ sound-crossing times without; 2) clouds can develop Jeans-instable regions internally even though they are initially below Jeans mass; 3) all clouds lose mass by ram pressure and Kelvin-Helmholtz instability; 4) thermal conduction substantially lowers mass-loss rates, by this, extending the clouds' lifetimes, particularly, more than doubling the lifetime of low-mass clouds; 5) thermal conduction leads to continuous, filamentary stripping, while the removed gas is heated up quickly and mixes efficiently with the ambient \cgm; 6) without thermal conduction the removed gas consists of dense, cool, clumpy fragments; 7) thermal conduction might prevent C\hvcs{} from forming stars; 8) clouds decelerated by means of drag from the ambient \cgm{} form head-tail shapes and collapse after they reach velocities characteristic for intermediate-velocity clouds.

Conclusively, only sophisticated modelling of C\hvcs{} as non-homogeneous and non-isothermal clouds with thermal conduction and self-gravity explains observed morphologies and naturally leads to the suppression of star formation.
%leads to local gravitational collapse.
\end{abstract}

\begin{keywords}
Conduction --
Diffusion --
Hydrodynamics --
Methods: numerical --
ISM: clouds
\end{keywords}

%%%%%%%%%%%%%%%%%%%%%%%%%%%%%%%%%%%%%%%%%%%%%%%%%%

%%%%%%%%%%%%%%%%% BODY OF PAPER %%%%%%%%%%%%%%%%%%

\section{Introduction}\label{sec:intro}
The coexistence of only gas phases in the interstellar medium (\ism) can theoretically be well described by the three-phase medium introduced by \citet{77mckeeostriker} \citep[applied to the halo by][]{95wolfireetal}. Stratified, cool, dense, molecular gas and cold neutral matter (CNM, $T\lesssim 100~$K) are surrounded by a warm, neutral (WNM) to slightly ionized medium (WIM, $8,000\lesssim T/{\rm K}\lesssim 10^{4.2}$). A hot, dilute, and highly ionized medium (\him, $T\gtrsim 10^{5.5}~$K) provides the third thermally stable phase. Gas with temperatures in one of the two ranges between the three phases is thermally instable and separates into either the cooler or the hotter phase by thermal instability \citep[][]{65field}. These two instable phases provide continuous pressure transitions between the stable phases \citep[][]{95wolfireetal,12vazquezsemadeni}. The physical conditions assumed for the \him{} are observed in the circumgalactic medium (\cgm) in the halo of our Milky Way \citep[][]{17tumlinsonpeepleswerk}. Among the stable phases the WIM has the highest volume filling factor, which raises from $\sim 0.4$ in the galactic plane to $\sim 0.6$ at a galactic height of $z=1~$kpc \citep[][]{87morfillscholer} and contains more than $50~$per cent of the total mass of the \ism{} \citep[][]{05tielens}. However, more recent observations and simulations reveal a much more complex dynamical and thermal state of the \ism{} that cannot be explained within the simplified theoretical framework of \citet{77mckeeostriker}: stars, electromagnetic radiation, cosmic rays, dust, magnetic fields, self-gravity, and turbulence contribute to the overall energy budget and dynamics and hence affect gas phases in various states (molecular, atomic, ionized, excited). These mutual interactions lead to many open questions \citep[][]{05cox,07mckeeostriker}. Especially for the \cgm{} it is not yet clear which small-scale physics can be deduced from the observed multiphase ionization structure \citep[][]{17tumlinsonpeepleswerk}.

By connecting the high pressure of CNM, which manifests in the form of gravitationally bound clouds, with the low pressure of \cgm, very steep gradients in both density and temperature are established in the WIM. A steep gradient in temperature between clouds and the hot gas results in the onset of a heat flux towards the cool cloud. Analytical approaches propose the evaporation of clouds in a few dynamical times \citep[see, e.g.][]{82balbusmckee,84giuliani,07nagashimainutsukakoyama}. However, since interstellar clouds are observed ubiquitously, two scenarios are intuitive to assume: (1) The accretion rate of intergalactic gas by the Milky Way and other galaxies in the local Universe is sufficient for replenishment \citep[][]{09bland-hawthorn,12richter,16richter}. (2) The clouds are stabilized against thermal (evaporation) and dynamical (e.g. ram-pressure stripping) interactions by processes that lead to accretion of mass, prevention of mass loss, and suppression of hydrodynamical instabilities. \citet[][]{07vieserhensler1} figured out thermal conduction to play a key role in stabilizing interstellar clouds over reasonable evolutionary times. The numerical investigations of \citet{04pittardetal} have shown that evaporation is the most likely mechanism for mass-loading of spherical accretion flows from embedded clouds if the ratio of accreted to ejected mass is high. Such a situation is observed in galactic winds from SNe II in the dwarf starburst galaxy NGC 1569 \citep[][]{02martinkobulnickyheckman} and simulated by \citet[][]{07recchihensler}. Pioneering investigations on the effect of thermal conduction on cool clouds were performed by \citet{77cowiemckee} and \citet{77mckeecowie}, who applied a saturation formalism to gaseous clouds in order to correctly describe the heat flux in the presence of steep temperature gradients. However, they did it in a stationary fashion, neglecting time-dependent contributions to the energy equation like mass diffusion, evaporation, condensation, metallicity (which in turn acts on the local cooling and heating functions of the plasma), and surface instabilities. Complementary to that, \citet{90begelmanmckee} developed a time-dependent theory of two-phase media under consideration of gas heating and radiative cooling, and with mass exchange via condensation and evaporation.

The variety of interstellar clouds comprises the distinct class of high-velocity clouds \citep[\hvcs, see][]{01wakker,01richteretal,14faridanietal}, which have line-of-sight velocities that are incompatible with the rotation speed of our Galaxy at their respective position. Seen in the Local Standard of Rest (LSR), their velocities are mainly negative implying a net infall to the Galaxy. They are observed primarily at higher galactic latitudes \citep[$|b|\gtrsim 25^\circ$,][]{05westmeierbruenskerp2} and throughout the entire electromagnetic spectrum. \hvcs{} are multi-phase clouds, which have a core-halo structure in temperature, density, and ionization fraction, and show a sub-solar metallicity. Due to a lack of suitable foreground and background sources the determination of their distances is still a challenge. Distance brackets put \hvc{} complexes in the Milky Way at distances between $1$ and $30~$kpc from the Sun. However, quite different from the \hvc{} complexes are isolated and compact \hvcs{} (C\hvcs), which are not being resolved by a $0.5^\circ$ beam and having angular sizes less than $2^\circ$ of FWHM \citep[][]{99braunburton}. They distribute rather uniformly over the sky \citep[][]{99burtonbraun,00braunburton} and neither connect spatially nor kinematically to any other feature of neutral hydrogen down to a column density of $\sim 1.5\times 10^{18}~$cm$^{-2}$ \citep[][]{04vanwoerdenetal}.
This lack of association leads to the definition of a distinct class of \hvcs{} that may have originated under the same environmental conditions, share a similar evolutionary history, and are in the same physical state. Among C\hvcs{} one finds clouds with the highest LSR velocities. C\hvcs{} can be located several $100~$kpc away from the Sun \citep[][]{00braunburton}.

Another puzzle is the total lack of stellar objects associated with \hvcs{} \citep[e.g.][]{05siegeletal}. But \citet{14izumietal} report on star formation in Cloud 1 (one of the Digel Clouds) triggered by the impact of the close-by \hvc{} complex H onto the Galactic disk (see also the numerical study by \citet[][]{94lepineduvert} for star formation induced by infall of \hvcs{} on the Galactic disk). Indeed, from observations it can be concluded that all necessary conditions for star formation in \hvcs{} are fulfilled: there exist candidates with sufficient mass, molecular hydrogen and dust, and a variety of elements allowing them to cool radiatively over decades of temperatures. Many of them show emission in both H$\alpha$ and infrared, but however, no stellar counterpart could have ever been assigned to these clouds \citep[e.g.][]{03hoppschulteladbeckkerp2,03hoppschulteladbeckkerp,15starkbakerkannappan}. However, a recent analysis of the \emph{SECCO} survey assigns an identified stellar system (most likely a faint, nearly starless dwarf galaxy located in the Virgo Cluster) to the ultra compact \hvc{}~$274.68$+$74.70$-$123$ \citep[][]{15bellazzinietal}.  

In this paper we systematically explore the effects of various physical processes on the evolution of C\hvcs{} moving through the \cgm. We emphasize, that we do not model the large complexes of \hvcs. The effects considered include self-gravity (\S~\ref{subsec:selfgravity}), a variable metallicity affecting heating and cooling of the plasma (\S~\ref{subsec:metallicity}), dissociation of molecules and equilibrium ionization of atoms (\S~\ref{subsec:dission}), collisional heating and radiative cooling of plasma (\S~\ref{subsec:heatcool}), thermal conduction (\S~\ref{subsec:thermalcond}), and diffusion of matter (\S~\ref{subsec:massdiff}). The individual processes are successively taken into account and the values for both cloud metallicity and mass are varied. We compare our results on the one hand to the findings of \citet{07vieserhensler1}, who conducted equal simulations in 2D on an equidistant grid under exploitation of cylindrical symmetry, in order to confirm their results with a more sophisticated 3D AMR numerical approach. Likewise, the simulations performed by \citet[][]{09heitschputman}, \citet[][]{17armillottaetal}, and \citet[][]{20lietal} are discussed in context of the conclusions drawn from simplified modelling of clouds as homogeneous and isothermal. On the other hand, we compare to the analytic calculations of \citet{77cowiemckee}, whose approach solely focuses on classical and saturated mass loss due to evaporation and neither accounts for self-gravity nor cooling. The latter is taken into account in \citet{77mckeecowie}.

This paper is structured as follows: in \S~\ref{sec:physproc} we describe the physical processes being considered in our simulations. In \S~\ref{sec:nummethod} we briefly explain the applied numerical code and \S~\ref{sec:modelsetups} provides our approach for setting up the model clouds. The results of the simulations are presented in \S~\ref{sec:modelcomparisons} and are finally discussed in \S~\ref{sec:discussion}. The paper is summarized in \S~\ref{sec:sumcon}, where also the major conclusions are drawn.

\section{Physical Processes}\label{sec:physproc}
In our simulation suite we have covered physical processes that most likely take place in interstellar clouds. These processes are explained below. Radiative transport is neglected, because we assume the model clouds to be optically thin. Moreover, chemodynamics is taken into account on a simplified level, i.e. different species are not traced explicitly. Instead, the fractions of H$_2$, \ion{H}{i}, \ion{H}{ii}, He, electrons, and metals are calculated semi-analytically by using local fractions of dissociation and ionization, respectively.

\subsection{Hydrodynamics}\label{subsec:eulerequations}
The dynamics of inviscid fluids in the absence of magnetic fields is determined by the Euler Equations
\begin{eqnarray}
\frac{\partial\varrho}{\partial t}+\nabla(\varrho\bm{v})&=&0, \label{equ:simulations-2}\\
\frac{\partial\varrho\bm{v}}{\partial t}+\nabla(\varrho\bm{v}\bm{v})+\nabla P&=&\varrho\bm{g}, \label{equ:simulations-3}\\
\frac{\partial\varrho e_{\rm tot}}{\partial t}+\nabla\left[(\varrho e_{\rm tot}+P)\bm{v}\right]&=&\sum_{j=1}^NS_{\!j}.\label{equ:simulations-4}
\end{eqnarray}
They yield the fluid's density $\varrho$, velocity $\bm{v}$, and total energy $e_{\rm tot}$ upon coupled integration. The $S_{\!j}$'s in Equ.~(\ref{equ:simulations-4}) above are arbitrary source terms modifying the total energy of the fluid. By an equation of state,
\begin{equation}
P=(\gamma-1)\varrho e_{\rm int},\label{equ:simulations-1}
\end{equation}
the pressure $P$ of the fluid is derived and the system of Euler Equations can be closed. Hence, the state of the fluid is determined for each and every time. The inner energy $e_{\rm int}$ is a function of the mean molecular weight $\mu$ thus Equ.~(\ref{equ:simulations-1}) implicitly respects each species in molecular, atomic, and ionized form (see \S~\ref{subsec:dission}). For solar metallicity $\mu$ converges to $\sim 2.4$ in the CNM, to $\sim 1.2$ in the WIM, and to $\sim 0.6$ in the \cgm. As we deal with only low-density environments in the simulations the assumption of an ideal gas is legitimated. Hence, $\gamma=5/3$ in all simulations.

\subsection{Self-gravity}\label{subsec:selfgravity}
The effect of self-gravity for an arbitrary matter distribution $\varrho(\bm{r})$ is described by solving the well-known Poisson's Equation
\begin{equation}
\nabla^2\Phi(\bm{r})=4\pi G \varrho(\bm{r}),\label{equ:selfgravity-1}
\end{equation}
for the gravitational potential $\Phi(\bm{r})$ ($G$ denotes Newton's gravitational constant). Each mass element of $\varrho(\bm{r})$ contributes to $\Phi(\bm{r})$. If the total mass becomes a significant fraction of the (temperature-dependent) Jeans mass $M_{\rm J}$ for a particular $\varrho(\bm{r})$, self-gravity cannot be neglected as it relevantly affects the dynamics of all mass elements. However, it has been shown in \citet{19sanderhensler} that self-gravity has a measurable effect on the evolution of even clouds with masses well below their Jeans mass.

If self-gravity is stronger than the acceleration of gas in the opposite direction (e.g. due to evaporation or advection by a streaming \cgm), the net acceleration at the cloud surface points in the same direction like the density gradient. Hence, the condition for Rayleigh-Taylor instability (\rti) is not given. So, it is expected that self-gravity significantly suppresses RTI at the cloud surface \citep[see][]{93murrayetal} and the typical finger-like structures do not form.

According to \citet{93murrayetal} the minimum mass of self-gravitating clouds above which they are stable against Kelvin-Helmholtz instability (\khi) when moving with a certain speed $v_{\rm rel}$ through an ambient medium with density $\varrho_{\rm\cgm}$ reads
\begin{equation}
M_{\rm min}\approx\frac{\sqrt{6}\pi v_{\rm rel}^3}{\sqrt{\bar{\varrho}_{\rm cloud}}}\left(\frac{\varrho_{\rm\cgm}}{G\bar{\varrho}_{\rm cloud}}\right)^{3/2},\label{equ:initial-5}
\end{equation}
with $\bar{\varrho}_{\rm cloud}$ being the average mass density in the cloud.

To ensure that self-gravitating clouds, which are embedded in a medium with pressure $P_{\rm\cgm}$, are gravitationally stable initially, their total mass must be below their \emph{Bonnor-Ebert mass} \citep{55ebert,56bonnor}
\begin{equation}
\mmax=1.18\left(\frac{k_{\rm B}\bar{T}_{\rm cloud}}{G^3\bar{\mu}_{\rm cloud}}\right)^2P_{\rm\cgm}^{-1/2}.\label{equ:initial-3}
\end{equation}
Here, $\bar{T}_{\rm cloud}$ and $\bar{\mu}_{\rm cloud}$ are the average temperature and mean molecular weight in the cloud.

\subsection{Metallicity}\label{subsec:metallicity}
Observations indicate a common metallicity for C\hvcs{} of $0.1$ to $0.3\zsolar$ \citep{01wakker}. In order to study extreme cases we consider two different initial cloud metallicities of $Z/\zsolar =0.1$ and $Z/\zsolar =1.0$. We note that the \cgm{} has constant metallicity in all runs (see \S~\ref{subsec:cgm}). The ratio of the abundances of hydrogen and helium is assumed solar and so are the ratios of metal abundances.

\subsection{Dissociation and ionization of elements}\label{subsec:dission}
The simulations contain cool clouds and hot plasma, so a temperature range from a few hundreds up to several $10^6~$K is covered. For $T<900~$K molecules exist, but they successively dissociate and finally become totally ionized with rising temperature. In our simulations we semi-analytically calculate both dissociation and ionization for hydrogen in thermodynamic equilibrium. The fraction of \ion{H}{ii} is then related to He and metals. We note that we do not trace single species but rather have one fluid for which the composition is calculated based upon a given metallicity. The fractions of dissociation and ionization affect the magnitudes of cooling rates. See appendix \ref{app:dission} for further details.

\subsection{Heating and cooling of plasma}\label{subsec:heatcool}
The rates of heating $\Gamma=\Gamma_0n$ and cooling $\Lambda=\Lambda_0n^2$ of the plasma are calculated semi-analytically by applying cooling laws from literature: for $T<900~$K the cooling curve from \citet{85falgaronepuget} is used, which considers molecular line cooling. For $900<T/{\rm K}<10^4$ the cooling function from \citet{72dalgarnomccray} is applied, and for temperatures above $10^4~$K the plasma cools according to the rates in \citet{89boehringerhensler}. Both $\Gamma$ and $\Lambda$ depend on the local values of metallicity, temperature, and density. These local variables are calculated in each time-step. The different cooling functions assume a solar ratio of metals.

The plasma is heated due to the photoelectric effect on dust particles \citep{06weingartnerdrainebarr}, ionization by UV radiation, by X-rays, and by cosmic rays \citep[][]{03wolfireetal}, thermalization of turbulent motions, and condensation of molecular hydrogen on dust particles \citep[both in][]{10tielens}.

\subsection{Thermal conduction}\label{subsec:thermalcond}
In the presence of large temperature gradients the classical description of heat flux given by \citet{62spitzer},
\begin{equation}
\bm{q}_{\rm class}=-\kappa_{\rm class}\nabla T,\label{equ:satcond-1}
\end{equation}
with the coefficient for thermal conduction $\kappa_{\rm class}=1.84\times 10^{-5}T^{5/2}/\ln(\Omega)$, and the Coulomb logarithm $\ln(\Omega)=29.7+\ln[T/(10^6\sqrt{n_{\rm e}})]$, yields a too high amount of energy being conducted by electrons \citep[e.g.][]{84campbell}. Instead, a \emph{saturated heat flux} $\bm{q}_{\rm sat}$ \citep[][]{77cowiemckee,77mckeecowie} accounts for a finite reservoir of electrons and thus converges to a maximum value if all electrons are conducting heat and irrespective of the steepness of temperature gradient. The transition from classical to saturated thermal conduction depends on both the mean free path of electrons in the \cgm,
\begin{equation}
\lel=t_{\rm e}\sqrt{\frac{3k_{\rm B}\tcgm}{m_{\rm e}}},\label{equ:parameter-2}
\end{equation}
where the electron-electron equipartition time \citep{62spitzer} reads
\begin{equation}
t_{\rm e}=\frac{3\sqrt{m_{\rm e}}(k_{\rm B}\tcgm)^{3/2}}{4\sqrt{\pi}n_{\rm e}e^4\ln(\Omega)},\label{equ:parameter-3}
\end{equation}
and the temperature scale height $\LT=T/|\nabla T|$ of the \cgm, which provides the typical spatial scale of temperature variations. It has been experimentally verified by \citet{80graykilkenny}, that
\begin{equation}
\bm{q}=\left\{
\begin{array}{ll}
\bm{q}_{\rm class} & ,\;\lel/\LT\lesssim 2\times 10^{-3} \\
\bm{q}_{\rm sat} & ,\;{\rm else}
\end{array} .
\right.\label{equ:satcond-2}
\end{equation}
\citet{77cowiemckee} suggest the form
\begin{equation}
q_{\rm sat}\equiv\left|\bm{q}_{\rm sat}\right|=5\Phi\varrho \cs^3,\label{equ:satcond-9}
\end{equation}
with a factor $\Phi$ accounting for uncertainties specific to the approximation of flux-limited diffusion and the effect of magnetic fields. It holds $\Phi=1.1$ if both ion and electron temperatures are equal. We use $\Phi=1$ throughout all of the simulations.

To ensure continuity at the transition between $\bm{q}_{\rm class}$ and $\bm{q}_{\rm sat}$ a proper representation for the heat flux must be chosen. Within this work the expression of \citet{92slavincox}
\begin{equation}
\bm{q}_{\rm eff}=\bm{q}_{\rm sat}\left(1-\exp\left\{-\frac{|\bm{q}_{\rm class}|}{|\bm{q}_{\rm sat}|}\right\}\right)\label{equ:satcond-3}
\end{equation}
is used. It correctly converges to either of the heat fluxes in Equ.~(\ref{equ:satcond-2}) for very small or very large temperature gradients, respectively.

The conductivity according to the effective heat flux~(\ref{equ:satcond-3}) reads
\begin{equation}
\kappa_{\rm eff}=\kappa_{\rm sat}\left(1-\exp\left\{-\frac{\kappa_{\rm class}}{\kappa_{\rm sat}}\right\}\right),\label{equ:satcond-5}
\end{equation}
with $\kappa_{\rm sat}=5\varrho \cs^3/\left|\nabla T\right|$ relating to the coefficient of thermal diffusion via $\alpha=\kappa_{\rm eff}/(\varrho c_V)$, where $c_V=\partial e_{\rm in}/\partial T$ is the isochoric heat capacity.

We follow the procedure described in \citet{07vieserhensler2} for calculating the temperature update according to saturated thermal conduction. From the new temperature distribution an adjusted heat flux 
\begin{equation}
\bm{q}_{\rm eff}=-\kappa_{\rm eff}\nabla T\label{equ:satcond-7}
\end{equation}
finally accounts for the change in energy by thermal conduction. The code test for only thermal conduction given by Equ.~(\ref{equ:satcond-7}) is shown in appendix \ref{app:codetest}.

If the wavelengths of temperature fluctuations in the ambient \cgm{} are greater than the Field length \citep[][]{90begelmanmckee}, 
\begin{equation}
\lf=\sqrt{\frac{\kappa_{\rm\cgm} \tcgm}{n^2\Lambda_{\rm max}}},\label{equ:resolution-1}
\end{equation}
these fluctuations are able to grow and lead to thermal instability ($\kappa_{\rm\cgm}$ and $\tcgm$ are the conductivity and temperature in the \cgm, $n$ is the local particle density and the maximum cooling rate reads $\Lambda_{\rm max}=\max\{\Lambda_0,\Gamma_0/n\}$). Thus, if $\rcloud\gg\lf$, the thermal content of the cloud is dominated by external heating and radiative cooling whereas for $\rcloud\ll\lf$ thermal conduction governs.

\begin{figure}
\centering
\includegraphics[width=.45\textwidth]{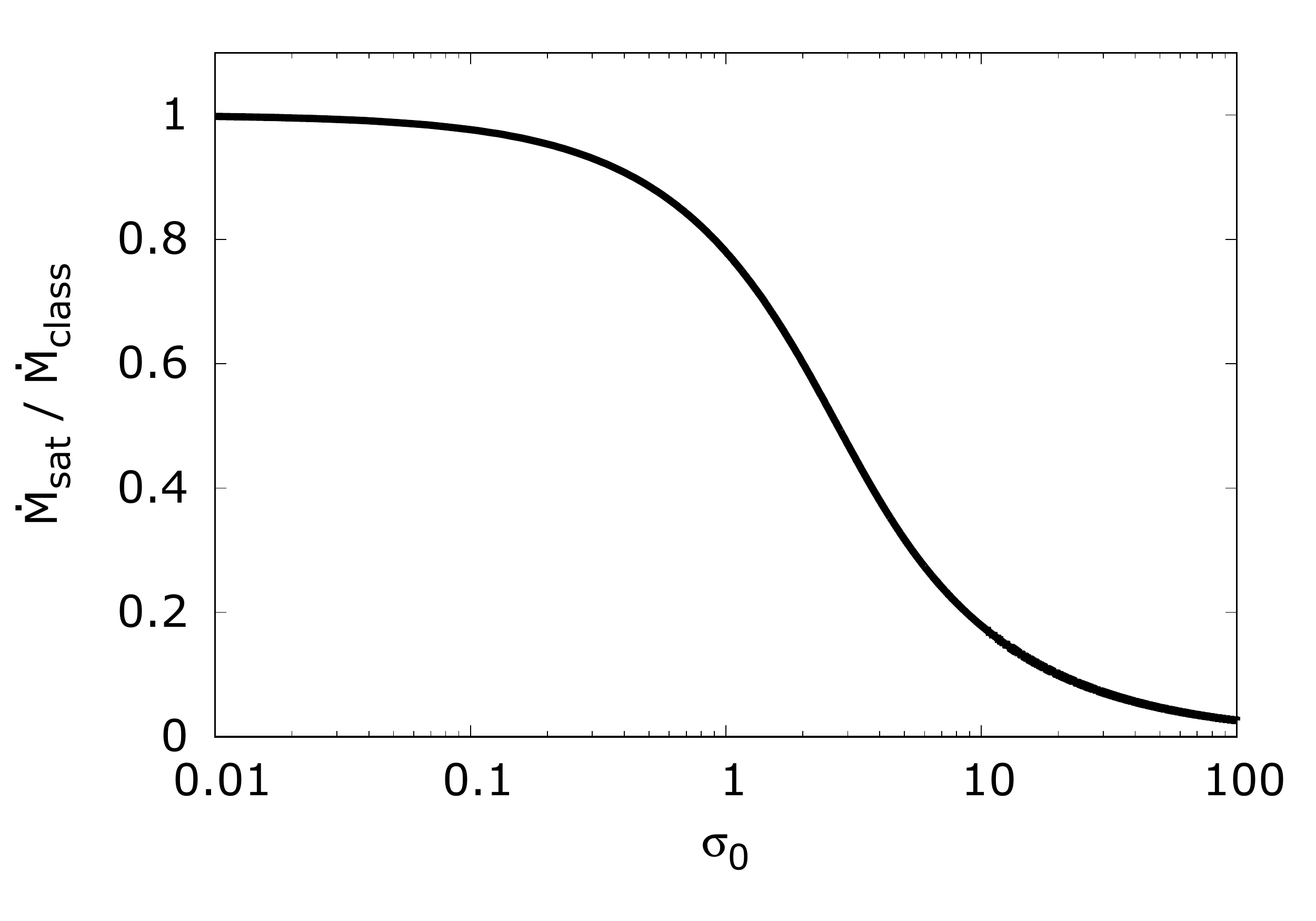}
\caption{Ratio of saturated to classical mass-transportation rate \citep[see][]{93daltonbalbus}.}
\label{fig:parameter-1}
\end{figure}

\citet{77cowiemckee} introduced a global saturation parameter
\begin{equation}
\sigma_0=\frac{\left(\tcgm/1.54\times 10^7{\rm [K]}\right)^2}{\ncgm\Phi \rcloud{\rm [pc]}},\label{equ:parameter-1}
\end{equation}
which can be used equivalently to Equ.~(\ref{equ:satcond-2}) to discriminate between classical ($\sigma_0<1$) and saturated ($\sigma_0>1$) thermal conduction. Here, $\rcloud$ is given in terms of parsec, and $\Phi$ is defined in Equ.~(\ref{equ:satcond-9}). The mass-loss rate by evaporation is considerably reduced with increasing $\sigma_0$ (Fig.~\ref{fig:parameter-1}). For $\sigma_0\gg 1$ also viscous heating becomes relevant \citep[][]{82balbusmckee}.
Theoretically, gas should evaporate from clouds for $\lel\lesssim \rcloud\ll\lf$ and condense onto them if $\rcloud\gtrsim\lf\gg\lel$. \citet{07vieserhensler2} have shown that the latter condition for condensation is already fulfilled for $\rcloud\gtrsim 0.24\lf$.

In order to correctly resolve $\lf$ in simulations, \citet{04koyamainutsuka,09gressel} suggest a resolution of $\lf$ by at least three grid cells. Otherwise, results become resolution dependent. In our simulations, $\lf$ is resolved by $158~$cells in the massive model clouds and $130~$cells in the low-mass clouds.

\subsection{Mass diffusion}\label{subsec:massdiff}
The simulations respect the transport of matter by diffusion using the formulation of the Chapman-Enskog expansion. This is a valid description at low gas densities \citep{60birdstewartlightfoot}. The mass diffusion flow reads
\begin{equation}
\bm{j}=-D\nabla\varrho,\label{equ:massdiff-1}
\end{equation}
with $D$ being the coefficient of mass diffusion. Assuming one species of gas,
\begin{equation}
D=0.0018583\frac{\sqrt{2T^3/\mu_{\rm H}}}{P\sigma_{\rm H}^2\Omega_{\rm int}}\label{equ:massdiff-2}
\end{equation}
according to the Chapman-Enskog expansion. Here, $\sigma_{\rm H}$ and $\mu_{\rm H}$ denote the cross section and molecular weight, respectively, of hydrogen. The collision integral is given by
\begin{eqnarray}
\Omega_{\rm int}&=&\frac{1.06036}{T_*^{0.15610}} + \frac{0.19300}{\exp\{0.47635T_*\}} \nonumber \\
&&+ \frac{1.03587}{\exp\{1.52996T_*\}} + \frac{1.76474}{\exp\{3.89411T_*\}},\label{equ:massdiff-3}
\end{eqnarray}
where $T_*=T/(E_{\rm LJ}/k)$ and $E_{\rm LJ}$ is the depth of the Lennard-Jones potential of hydrogen.

\subsection{Magnetic fields}\label{subsec:magfields}
We explicitly neglect magnetic fields, because so far only four rather imprecise measurements of magnetic field strengths in \hvcs{} without any clue on orientation are available. \citet{91kazestrolandcrutcher} searched for the Zeeman effect in four \ion{H}{i} \hvcs, where they measured a magnetic field $|\bm{B}|=(-11.4\pm 2.4)~\mu$G in only \hvc~$132$+$23$-$212$. More recently, \citet{10mccluregriffithsetal} reported on the strength for the line-of-sight component of the coherent magnetic field of $\gtrsim 6~\mu$G in the \hvc{} associated with the Leading Arm of the Magellanic System. \citet[][]{13hilletal} and \citet[][]{19bettietal} detected a magnetic field with strength $\gtrsim 8~\mu$G and $\gtrsim 5~\mu$G, respectively, associated with the Smith Cloud. Hence, magnetic fields of \hvcs{} are weakly constrained by observations in orientation and strength. Another plausible cause for Faraday rotation can be the compressed magnetic field in the \cgm{} in front of infalling \hvcs{} \citep[][]{11jelinekhensler}. In this case, the source of Faraday rotation does not belong to the cloud. So, neglecting magnetic fields in our models does not contradict the observations.

It further has been shown theoretically by \citet{19sanderhensler} that the effective heat flux, i.e. integrated over the cloud surface, is reduced to only $68~$per cent of its undisturbed value by a strong magnetic dipole field. Thus, neglecting magnetic fields in our models is theoretically justified with respect to thermal conduction.

\section{Numerical methods}\label{sec:nummethod}
\subsection{The code}\label{subsec:code}
We carry out our simulations by using the publicly available code {\sc Flash}\footnote{See \href{http://flash.uchicago.edu/site/flashcode/}{http://flash.uchicago.edu/site/flashcode/}} \citep[][]{00fryxelletal,09dubeyetal}. It is a Eulerian hydrodynamics code, by which the (inviscid) Euler Equations~(\ref{equ:simulations-2})~to~(\ref{equ:simulations-4}) are integrated on a Cartesian grid by means of the piecewise parabolic method \citep[PPM,][]{84collelawoodward,84woodwardcolella}.

Our simulations are performed in full $3$ dimensions with outflow boundary conditions for the hydrodynamics at all sides of the computational domain, i.e. energy and mass can leave the domain, but are not allowed to return. We aim at investigating in detail the evolution of cloud structures thus the adaptive mesh refinement (AMR) property of {\sc Flash} is extensively utilized. The local numerical resolution is increased (decreased) based on the local density gradient by subsequently doubling (halvening) the number of blocks per dimension. One block consists of $8^3$ grid cells. The side length of the computational cube in all of our models is $340~$pc and the finest spatial resolution $\Delta x=1.3~$pc with six levels of AMR.

We use the Multigrid solver provided by {\sc Flash} for solving Poisson's Equation \citep[for an introduction to multilevel methods see][]{03zumbusch} together with isolated boundary conditions. By this, the potential field in the cells at the boundary of the computational domain can be calculated by taking appropriate density values from so-called ghost cells surrounding the entire domain. The domain of integration is extended by four layers of ghost cells at each face, which get their physical values by copying them from the outermost cells of the domain. The same holds for each and every block the computational domain is subsequently divided into.

\subsection{Time-steps}\label{subsec:timesteps}
In multiphysics simulations one is restricted by the time-step of the fastest process when integrating Equs.~(\ref{equ:simulations-2})~to~(\ref{equ:simulations-4}). Conservatively, integration is performed on the minimum value of all such time-steps in the entire computational domain. So, the wallclock times of simulations can increase enormously. 

First, plasma cooling evolves on a time scale
\begin{equation}
\Delta t_{\rm cool}=\frac{\varrho e_{\rm int}}{|\Gamma - \Lambda|+\varepsilon},\label{equ:timesteps-1}
\end{equation}
where $\varepsilon>0$ is a small number that prevents the code from dividing by zero at thermal equilibrium in a grid cell. In regions dominated by cooling $\Delta t_{\rm cool}$ can be very small.

Differently, thermal conduction is determined by the conduction time 
\begin{equation}
\Delta t_{\rm cond}=\frac{\min\{\Delta x^2,\Delta y^2,\Delta z^2\}}{\max\{\alpha\}},\label{equ:timesteps-3}
\end{equation}
where $\alpha=\kappa_{\rm eff}/(\varrho c_V)$ is the coefficient of thermal diffusion, $c_V=\partial e_{\rm in}/\partial T$ is the isochoric heat capacity, and the $\Delta$'s are the spatial resolutions in $x$-, $y$- and $z$-direction. Near cloud boundaries, the numerical grid is fine-grained and $\alpha$ becomes very high thus $\Delta t_{\rm cond}$ is remarkably scaled down.

It thus appears to be very likely in our model setups that both $\Delta t_{\rm cool}$ and $\Delta t_{\rm cond}$ are much smaller than
\begin{equation}
\Delta t_{\rm CFL}=C_{\rm CFL}\frac{\min\left\{\Delta x,\Delta y,\Delta z\right\}}{\max\left\{|\bm{v}|+\cs\right\}},\label{equ:timesteps-2}
\end{equation}
which is the numerically allowed maximum time-step according to the Courant-Friedrichs-Lewy condition \citep{28courantfriedrichslewy} and ensures a numerically stable solution of Equs.~(\ref{equ:simulations-2}) to (\ref{equ:simulations-4}). Depending on the fluid speed $|\bm{v}|$ we set the safety factor $C_{\rm CFL}=0.1$, $0.2$, or $0.4$ in our simulations.

To circumvent this numerical restriction we apply implicit integration methods. To update the energy based on heating and cooling (\S~\ref{subsec:heatcool}) we apply an implicit bisection method \citep[cf.][]{11engelnmuellgesniederdrenkwodicka}. Thermal conduction (\S~\ref{subsec:thermalcond}) is numerically addressed by the semi-implicit Crank-Nicolson method \citep[][]{47cranknicolsonhartree}, which we have implemented in the {\sc Flash} code.

Hence we are neither restricted by $\Delta t_{\rm cool}$ nor $\Delta t_{\rm cond}$ and the total energy density $\varrho e_{\rm tot}$ can be updated according to
\begin{equation}
\frac{\Delta \varrho e_{\rm tot}}{\Delta t}+\nabla(\varrho e_{\rm tot}\bm{v})=-\nabla(P\bm{v})+\varrho\bm{v}\bm{g}+\Gamma-\Lambda-\nabla\bm{q}_{\rm eff},\label{equ:timesteps-4}
\end{equation}
where $\Delta t$ is given by Equ.~(\ref{equ:timesteps-2}).

\subsection{Slope limiter}\label{subsec:slopelimiter}
Another numerical complicacy is due to very steep temperature gradients across cloud surfaces. The heat flux~(\ref{equ:satcond-7}) can thus be very high in the vicinity of regions of nearly zero heat flux. By this, artificial numerical oscillations can be produced near steep temperature gradients when integrating Equs.~(\ref{equ:simulations-2})~to~(\ref{equ:simulations-4}). In order to subsequently limit these oscillations we use a minmod (or: minbee) slope limiter to smooth temperature gradients. The minmod slope limiter preserves monotonicity of the employed upwind scheme thus ensuring diminishing of total variation \citep{09toro} of the solution even near discontinuities. By definition, the slope limiter becomes significant only around large temperature gradients. In regions of equal temperature it is simply equal to zero. 

\section{Model setups}\label{sec:modelsetups}
\subsection{Initial cloud profiles}\label{subsec:initialprofiles}
\begin{figure}
\centering
\includegraphics[width=.5\textwidth]{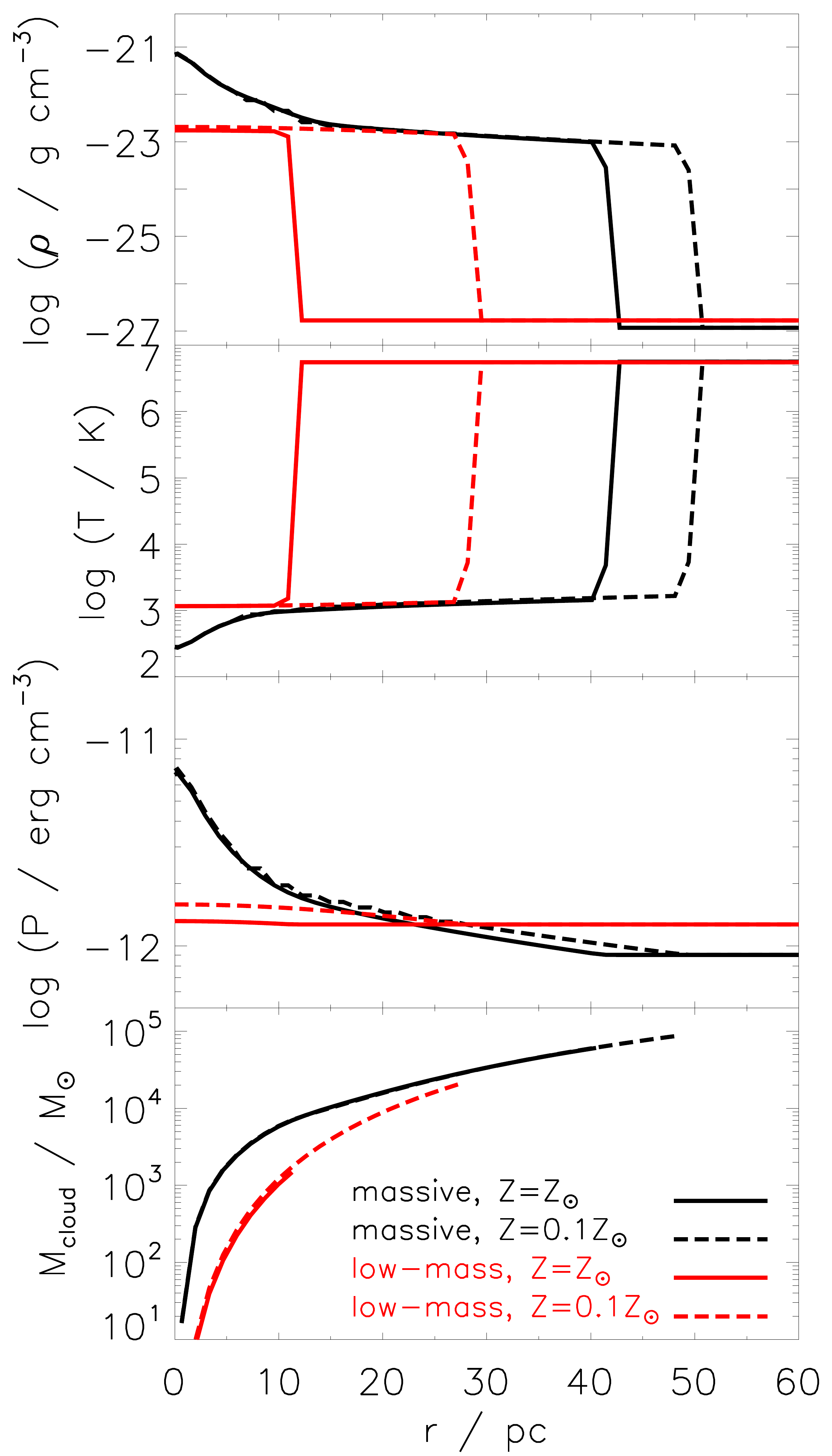}
\caption{\emph{From top to bottom:} Initial radial profiles of density, temperature, pressure, and mass for all model clouds.}
\label{fig:initial-1}
\end{figure}

The gas in the clouds is stratified hydrostatically, i.e. the condition for hydrostatic equilibrium
\begin{equation}
-\frac{\partial P}{\partial r}=\varrho\frac{GM(r)}{r^2}\label{equ:initial-1}
\end{equation}
must hold. In addition, the cloud is in thermal equilibrium at each and every point, i.e. the energy input by external heating is compensated by radiative cooling (cf. \S~\ref{subsec:heatcool}). We realize this condition by adapting the density to a certain temperature value, such that $\Lambda n^2=\Gamma n$. By that, matter density becomes a function of local temperature, $\varrho=\varrho(T)$. By substituting $P$ from Equ.~(\ref{equ:simulations-1}) into Equ.~(\ref{equ:initial-1}) we derive the governing equation for the radial temperature profile of the cloud,
\begin{equation}
\frac{\partial T}{\partial r}=-\varrho\left(T(r)\right)\frac{GM(r)}{r^2}\left\{\frac{\partial}{\partial T}\left[\frac{\varrho\left(T(r)\right)k_{\rm B}T(r)}{\mu\left(\varrho,T(r)\right)}\right]\right\}^{-1},\label{equ:initial-2}
\end{equation}
where $\mu$ is the mean molecular weight. In order to solve Equ.~(\ref{equ:initial-2}) the central cloud temperature $\tcentre$ and the temperature and particle density of the \cgm, $\tcgm$ and $\ncgm$, must be provided as boundary conditions (Table~\ref{table:initial-3} shows our chosen values). Then, Equ.~(\ref{equ:initial-2}) is integrated outwards with the density being assigned according to thermal equilibrium. The radial integration is performed until the pressure of the cloud matches that of the \cgm. By that, the cloud radius $\rcloud$ is a natural consequence of pressure equilibrium. The radial profiles of the initial configurations for all model clouds are given in Fig.~\ref{fig:initial-1}.
\begin{table}
\caption{Boundary conditions for obtaining the initial cloud models by integrating Equ.~(\ref{equ:initial-2}).}
\label{table:initial-3}
\centering
\begin{threeparttable}
\begin{tabular}{l c c}  
\hline
Parameter & massive clouds (M) & low-mass clouds (L) \\
\hline
$\tcentre~$/ K & $380$ & $1,150$ \\
$\tcgm$ / $10^6~$K & $5.6$ & $5.5$ \\
$\ncgm$ / $10^{-3}~$cm$^{-3}$ & $0.7$ & $1.0$ \\
\hline
\end{tabular}
\end{threeparttable}
\end{table}
At constant \cgm{} temperature the particle density of \cgm{} is the only parameter effecting the ambient pressure. It thus controls the cloud radius when solving Equ.~(\ref{equ:initial-2}). By inspecting the third plot in Fig.~\ref{fig:initial-1} one observes a higher ambient pressure in the model setups for the low-mass clouds due to a higher ambient particle density. Thus, pressure balance is reached at a smaller radius.

\subsection{Cloud masses}\label{subsec:masses}
The solution of Equ.~(\ref{equ:initial-2}) is mapped to the numerical grid of {\sc Flash} code, which consists of cells of equal size (adaptive mesh refinement is applied after the initial conditions are mapped to the grid). Due to the constraint of thermal equilibrium the density is simultaneously provided, such that the mass of the cloud is given by adding together the masses of all grid cells inside the cloud radius (lower plot in Fig.~\ref{fig:initial-1}).

In order to obtain a cloud that is dynamically stable initially and conducts thermal energy in a saturated manner it must be constrained by the conditions
\begin{equation}
\sigma_0 > 1\quad ,\quad\rcloud\lesssim\lf\quad ,\quad M_{\rm cloud}\in [M_{\rm min},\mmax],\label{equ:initial-6}
\end{equation}
whereas the \cgm{} must satisfy
\begin{equation}
\lel\gtrsim \LT.\label{equ:initial-7}
\end{equation}
Table~\ref{table:initial-2} contains the realisations of our setups in the parameter sub-spaces~(\ref{equ:initial-6})~and~(\ref{equ:initial-7}). The values for $\mcloud$ are listed in Table~\ref{table:Initial_Setup_2}.
\begin{table}
\caption{Initial parameters characterising the simulation models with respect to constraints (\ref{equ:initial-6})~and~(\ref{equ:initial-7}). The values are specific to clouds with $Z/\zsolar =0.1$ (in brackets for $Z/\zsolar =1.0$).}
\label{table:initial-2}
\centering
\begin{threeparttable}
\begin{tabular}{l c c}  
\hline
Parameter & massive clouds & low-mass clouds \\
\hline
$\sigma_0$ & 3.8 (4.6) & 4.6 (11.2) \\
$\mcloud/M_{\rm min}$ & $>3$ & $<0.8$ \\
$\mcloud/\mmax$ & 0.25 (0.23) & 0.08 ($\lesssim 0.01$) \\
$\rcloud/\lf$\tnote{(a)} & $\leq 0.23$ (0.20) & 0.16 (0.07) \\
$\lel/\LT$ & 1.05 (1.21) & 0.59 (0.68) \\
\hline
\end{tabular}
\begin{tablenotes}\footnotesize
\item[(a)] We point out that $\lf$ is not defined for clouds in which thermal conduction is not considered.
\end{tablenotes}
\end{threeparttable}
\end{table}

\subsection{Ambient \cgm}\label{subsec:cgm}
Our current understanding of the \cgm{} only allows hints to the range of density and temperature inhomogeneities, which are not fully understood \citep[as e.g. heating of the intra-cluster medium (ICM) by the motion of galaxies, see][]{90justetal} and must act on large scales. An example is the suppression of the growth of filaments in the ICM above a critical length scale $l_{\rm crit}$, derived by \citet[][]{04nipotibinney}. Inserting our cloud parameters into their equ.~(6) yields $l_{\rm crit}=26~$kpc (massive clouds) and $l_{\rm crit}=15~$kpc (low-mass clouds). Our fastest model clouds are M1c\_vel and L1b\_vel. So, within $60~$Myr they travel distances of $20~$kpc (M1c\_vel) and $15~$kpc (L1b\_vel), which do not exceed $l_{\rm crit}$ in either case. Thus, we examine constant \cgm{} conditions (cf. Table~\ref{table:initial-3}). Assuming a galactic stratification of ionized gas \citep[see equ.~(6) in][]{08gaensleretal} with a scale height of $2.6\pm 0.4~$kpc \citep[derived for the distribution of warm gas in the \cgm{} based on observations of \ion{Si}{iv} and \ion{O}{vi} by][]{19qubregman} our model clouds can be roughly placed at a galactic height of $\sim 100~$kpc.

Although the \cgm{} metallicity is uncertain around a range of $0.3$ to $0.4\zsolar$ we set the metallicity of the \cgm{} to be solar in order to better distinguish the amount of \cgm{} accreted by the cloud and the location of mixed C\hvc{} gas. For densities and temperatures we adapt for the \cgm{} the metallicity does not play a crucial role, because cooling is mainly due to Bremsstrahlung. However, for mixing phases at $\lesssim 10^5~$K metal lines of various ionization stages determine the cooling strength such that metallicity matters there.

\subsection{Subsonic flow}\label{subsec:flowint}
Our model C\hvcs{} are subject to a subsonic flow of hot gas representing the motion of the clouds through the \cgm{} of the Milky Way. By that, we cover the velocity range between $83$ and $333\kms$ (Table~\ref{table:Initial_Setup_2}). 

The flow is initially set up as a potential flow around the cloud, which is valid for subsonic flows \citep{66landaulifshitz}. Since the cloud is initially a sphere, the velocity field of the flow at a distance $|\bm{d}|$ from the cloud is given by
\begin{equation}
\bm{v}(\bm{d})=\left\{
\begin{array}{ll}
\bm{v}_{\rm \cgm}+\frac{\rcloud^3}{2|\bm{d}|^3}\left[\bm{v}_{\rm \cgm}-3\bm{n}(\bm{v}_{\rm \cgm}\cdot\bm{n})\right] & ,|\bm{d}|>\rcloud \\
0 & ,|\bm{d}|\leq \rcloud
\end{array},
\right.\label{equ:flow-1}
\end{equation}
with $\bm{n}$ being the unit vector pointing in the direction of $\bm{d}$. Keep in mind that a potential flow provides a laminar initial condition. In most of our simulation models the velocity of the flow is constant. However, in two of the models the velocity of the C\hvc{} relative to \cgm{} decreases due to the drag force of the \cgm. The decreased cloud velocity at time $t_{i+1}$ is calculated by subtracting
\begin{equation}
{\rm d} v=\frac{C_D}{2}\frac{A_{\rm cloud}}{\mcloud}\rhocgm v^2\left(t_i\right)\Delta t\label{equ:M1dec-1}
\end{equation}
from the cloud velocity at time $t_i$, where $A_{\rm cloud}$ and $\mcloud$ are the effective area and the mass of the cloud, respectively, and $\Delta t$ is the current integration time-step. The drag coefficient $C_D$ is calculated according to \citet[]{76henderson} and reads
\begin{equation}
C_D=C_1+C_2+C_3,\label{equ:M1dec-2}
\end{equation}
where
\begin{eqnarray}
C_1&=&24\left\{Re+v_{\rm mol}\left[4.33+\frac{3.65-1.53\tcloud/\tcgm}{1+0.353Re+0.48\sqrt{Re}}\right.\right.\nonumber\\
&&\left.\left.\times\exp\left\{-0.247\frac{Re}{v_{\rm mol}}\right\}\right]\right\}^{-1}\label{equ:M1dec-3},\\
C_2&=&\exp\left\{-\frac{Ma}{2\sqrt{Re}}\right\}\frac{4.5+0.38\left(0.03Re+0.48\sqrt{Re}\right)}{1+0.03Re+0.48\sqrt{Re}}\nonumber\\
&&+0.1Ma^2+0.2Ma^8\label{equ:M1dec-4},\\
C_3&=&0.6v_{\rm mol}\left(1-\exp\left\{-\frac{Ma}{Re}\right\}\right).\label{equ:M1dec-5}
\end{eqnarray}
The Reynolds number is $Re=\rhocgm v\left(t_i\right)\rcloud/\eta$, with dynamic viscosity $\eta=0.1$, and the molecular speed ratio reads $v_{\rm mol}~=~Ma\sqrt{0.5\gamma}$, with $\gamma$ being the ratio of specific heats and $Ma$ is the Mach number of the flow. The drag coefficient is calculated for isothermal clouds in \citet[]{76henderson} so we set $\tcloud$ in Equ.~(\ref{equ:M1dec-3}) as the boundary temperature in our model clouds, since this is the temperature at interface to \cgm. Equs.~(\ref{equ:M1dec-2})~to~(\ref{equ:M1dec-5}) are valid expressions for subsonic flows.

\subsection{Model clouds}\label{subsec:modelclouds}
\begin{figure}
\centering
\includegraphics[width=.5\textwidth]{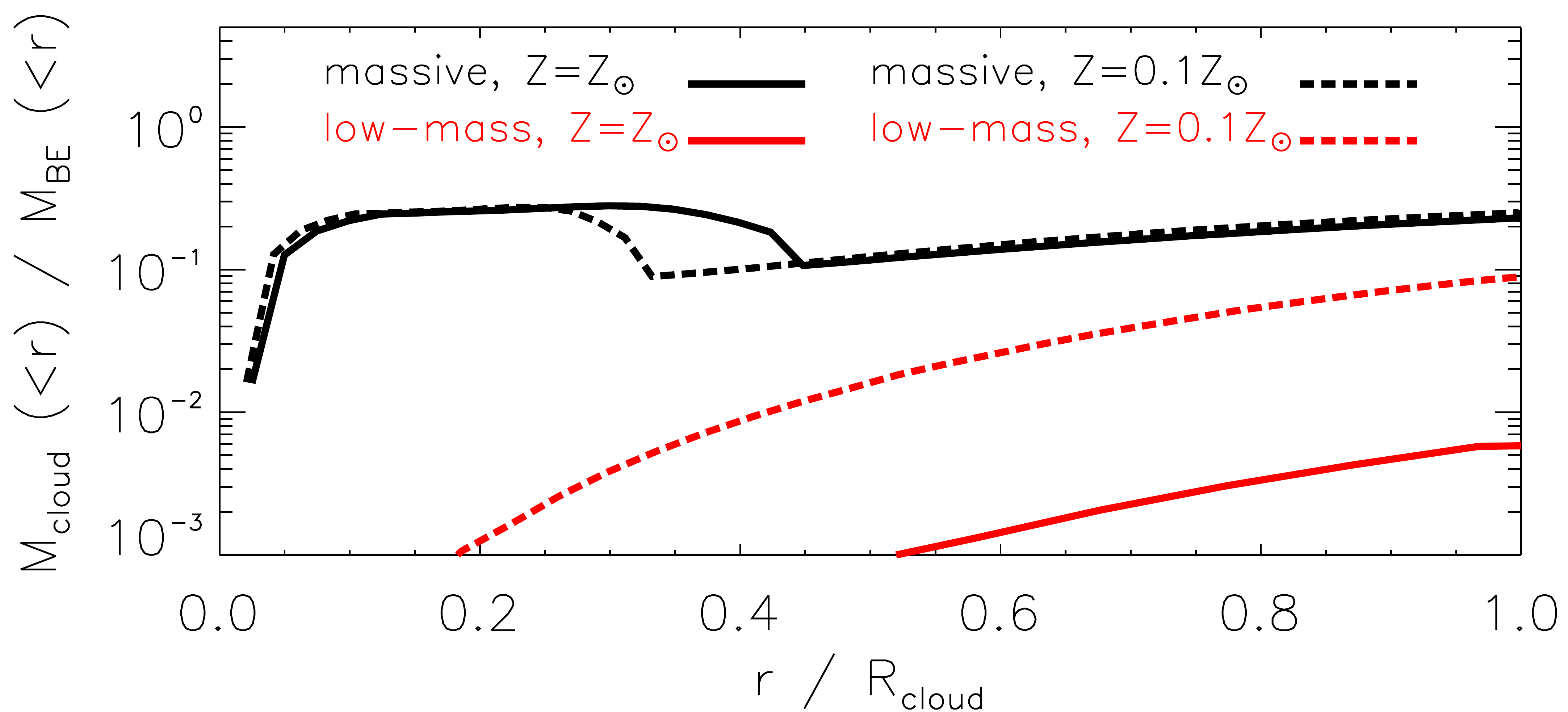}
\caption{Radial cloud-to-Bonnor-Ebert mass ratios for all model clouds. The x-axis is normalized to the respective radius of each cloud.}
\label{fig:initial-2}
\end{figure}

We have simulated and analysed models differing in the considered physical processes or in model parameters, which are key to their evolution. In Table~\ref{table:Initial_Setup_2} the models are listed. There are $14$ models in total allowing us to compare them pairwise and give clues on the effect of particular physics. All clouds are initially in both internal hydrostatic and thermal equilibrium as well as in pressure equilibrium with the ambient \cgm. All models take into account heating and radiative cooling of plasma. The cloud speed is constant in all but the models with decelerated clouds. The clouds are divided into the two groups of massive (M) and low-mass (L) clouds.
\begin{enumerate}
\item The massive clouds have a steep radial temperature profile which allows for a molecular phase in the centre and an extended atomic phase at larger radii. Only a small shell at the boundary consists of warm, slightly ionized gas. The density profile at all radii forms a distinct core-halo structure. All massive clouds are dynamically stable regarding $\mcloud>M_{\rm min}$, but have masses below their respective Bonnor-Ebert mass ($<0.3~\mmax$, see Fig.~\ref{fig:initial-2}), which is a non-linear function of radius as the massive clouds are not isothermal. To study the effect of physical simplicity we set up model M3\_hom\_tc\_sg, which is isothermal with $\tcloud=10^3~$K and homogeneous with $\rhocloud=1.5\times 10^{-23}~$g cm$^{-3}$. The sound speed in the \cgm{} has a value of $359\kms$.\item The low-mass clouds have a higher central temperature yielding a lower central density with respect to thermal equilibrium. Consequently, the central pressure is low and together with a higher value of ambient particle density a smaller radius is obtained. These clouds lack the molecular phase in their centres and they are only weakly gravitationally bound. According to the mass ratio $\mcloud/M_{\rm min}$ (Table~\ref{table:initial-2}) these clouds are expected to be disrupted. We can thus study if the life-time of clouds, which are disrupted anyway, is increased by thermal conduction. Likewise, their masses are below the Bonnor-Ebert mass ($<0.1~\mmax$, see Fig.~\ref{fig:initial-2}). As the low-mass clouds are almost isothermal, there is no such distinct feature in the radial profile of $\mmax$.
The sound speed in the \cgm{} has a value of $355\kms$.
\end{enumerate}
For both massive and low-mass clouds we assume two different metallicities to quantify the effect of metal-dependent radiative cooling: $Z/\zsolar =1.0$ is chosen in order to be comparable to the clouds investigated by \citet{07vieserhensler2} and $Z/\zsolar =0.1$ is more likely for \hvcs{} falling onto the Milky Way \citep[][]{01wakker}. Furthermore, the processes of thermal conduction and self-gravity are either considered in the models or not (cf. Table~\ref{table:Initial_Setup_2}) in order to study their respective impact on cloud evolution. Different cloud velocities scan the range of actually observed LSR velocities.

\subsection{Reference clouds}\label{subsec:referenceclouds}
In order to study the effects of different parameters and processes we define one respective reference cloud for each of the massive and low-mass models. Within the reference models the physical processes of self-gravity and thermal conduction are considered, their simulated velocities are typical for C\hvcs{}, and they have a metallicity of $0.1\zsolar$. These cloud models most resemble actually observed C\hvcs{} \citep[][]{12williamsetal}.
\begin{table*}
\caption{Simulated model clouds with initial values of radius (\emph{column 5}), mass (\emph{column 6}), metallicity (\emph{column 7}), speed (\emph{column 8}), and Mach number (\emph{column 9}). \emph{Columns 3} and \emph{4} show the physical processes that are either considered ($+$) or neglected ($-$).}              
\label{table:Initial_Setup_2}
\centering
\begin{threeparttable}
\begin{tabular}{c l c c c c c c c c}
\hline
Model group & Model & self-gravity & thermal conduction & $\rcloud/$pc & $\mcloud/10^4\msolar$ & $Z/\zsolar$ & $v_{\rm cloud}/\kms$ & Ma \\ 
\hline
\parbox[t]{2mm}{\multirow{8}{*}{\rotatebox[origin=c]{90}{massive clouds}}} & M0 (reference model) & $+$ & $+$ & $49$ & $9.0$ & $0.1$ & $167$ & $0.46$ \\
& M1\_tc & $+$ & $-$ & $49$ & $9.0$ & $0.1$ & $167$ & $0.46$ \\
& M1a\_vel & $+$ & $+$ & $49$ & $9.0$ & $0.1$ & $83$ & $0.23$ \\
& M1b\_vel & $+$ & $+$ & $49$ & $9.0$ & $0.1$ & $250$ & $0.69$ \\
& M1c\_vel & $+$ & $+$ & $49$ & $9.0$ & $0.1$ & $333$ & $0.93$ \\
& M1\_dec & $+$ & $+$ & $49$ & $9.0$ & $0.1$ & $167$\tnote{($*$)} & $0.46$\tnote{($*$)} \\
& M1\_me & $+$ & $+$ & $41$ & $6.3$ & $1.0$ & $167$ & $0.46$ \\
& M3\_hom\_tc\_sg & $-$ & $-$ & $41$ & $6.4$ & $1.0$ & $167$ & $0.46$ \\
\hline
\parbox[t]{2mm}{\multirow{6}{*}{\rotatebox[origin=c]{90}{low-mass clouds}}} & L0 (reference model) & $+$ & $+$ & $28$ & $2.2$ & $0.1$ & $165$ & $0.46$ \\
& L1\_tc & $+$ & $-$ & $28$ & $2.2$ & $0.1$ & $165$ & $0.46$ \\
& L1a\_vel & $+$ & $+$ & $28$ & $2.2$ & $0.1$ & $83$ & $0.23$ \\
& L1b\_vel & $+$ & $+$ & $28$ & $2.2$ & $0.1$ & $248$ & $0.69$ \\
& L1\_dec & $+$ & $+$ & $28$ & $2.2$ & $0.1$ & $165$\tnote{($*$)} & $0.46$\tnote{($*$)} \\
& L1\_me & $+$ & $+$ & $11$ & $0.15$ & $1.0$ & $165$ & $0.46$ \\
\hline
\end{tabular}
\begin{tablenotes}\footnotesize
\item[($*$)] Cloud speed is a function of time according to the drag force of ambient \cgm.
\end{tablenotes}
\end{threeparttable}
\end{table*}

\section{Model comparisons}\label{sec:modelcomparisons}
In the following we compare the simulated model clouds according to their differences as shown in Table~\ref{table:Initial_Setup_2}. Hence, the respective clouds are compared in terms of their masses, the effect of thermal conduction, their speed together with the effect of drag by the surrounding medium, and the strength of radiative cooling determined by the metallicity in the clouds. Finally, the effect of physical simplicity is described.

For the purpose of tracking physical quantities (e.g. velocity fields) and locating physical processes (e.g. \khi, gravitational collapse, mixing of gas) in the clouds we study slices through their density distributions. Integrated quantities, like column density, are discussed separately.

\subsection{Cloud mass}\label{subsec:mass}
\begin{figure}
\centering
\includegraphics[width=\linewidth]{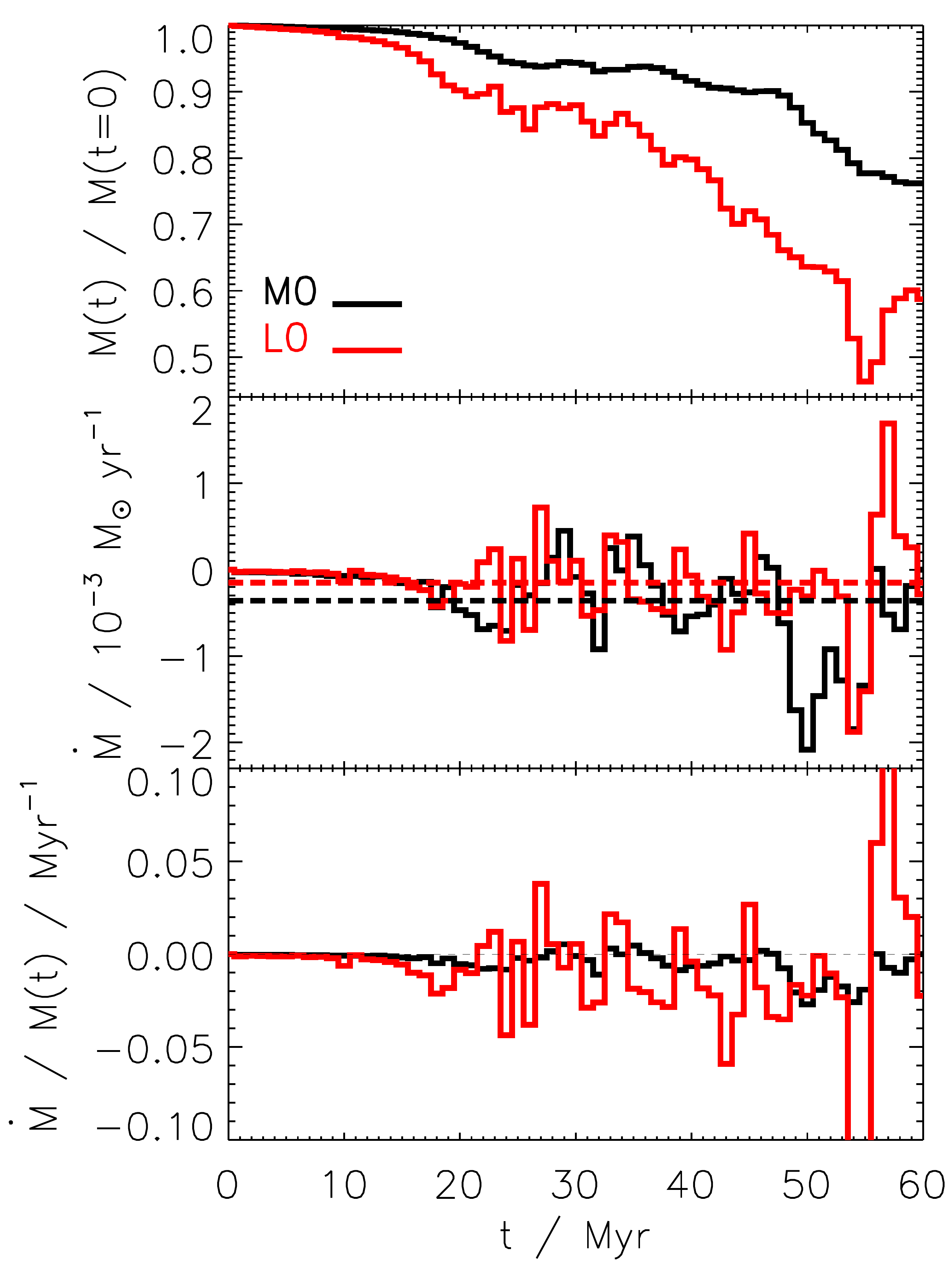}
\caption{Evolution of mass (\emph{upper plot}) for reference clouds. The corresponding mass-transportation rates (\emph{middle plot}) show both loss ($\dot{M}<0$) and gain ($\dot{M}>0$) of mass with the average values (\emph{dashed lines}) indicating a mass loss. The relative mass loss (\emph{lower plot}) is higher in low-mass clouds.}
\label{fig:mass-1}
\end{figure}

Both massive and low-mass clouds effectively lose mass during their passage through the \cgm{} with mean rates $\dot{M}$ of $-3.6\times 10^{-4}\msolar~$yr$^{-1}$ and $-1.6\times 10^{-4}\msolar~$yr$^{-1}$ (models M0 and L0, see Table \ref{table:massloss-2}). Low-mass clouds can lose up to $5~$per cent of their current mass per Myr after $20~$Myr and more than $10~$per cent after $50~$Myr (Fig.~\ref{fig:mass-1}, lower plot) while for massive clouds their mass loss is always less than $3$ per cent per Myr. After $60~$Myr of evolution there are only about $60~$per cent of initial mass left in the low-mass clouds while the massive clouds still contain $76~$per cent.

\begin{figure*}
\centering
\includegraphics[width=\textwidth]{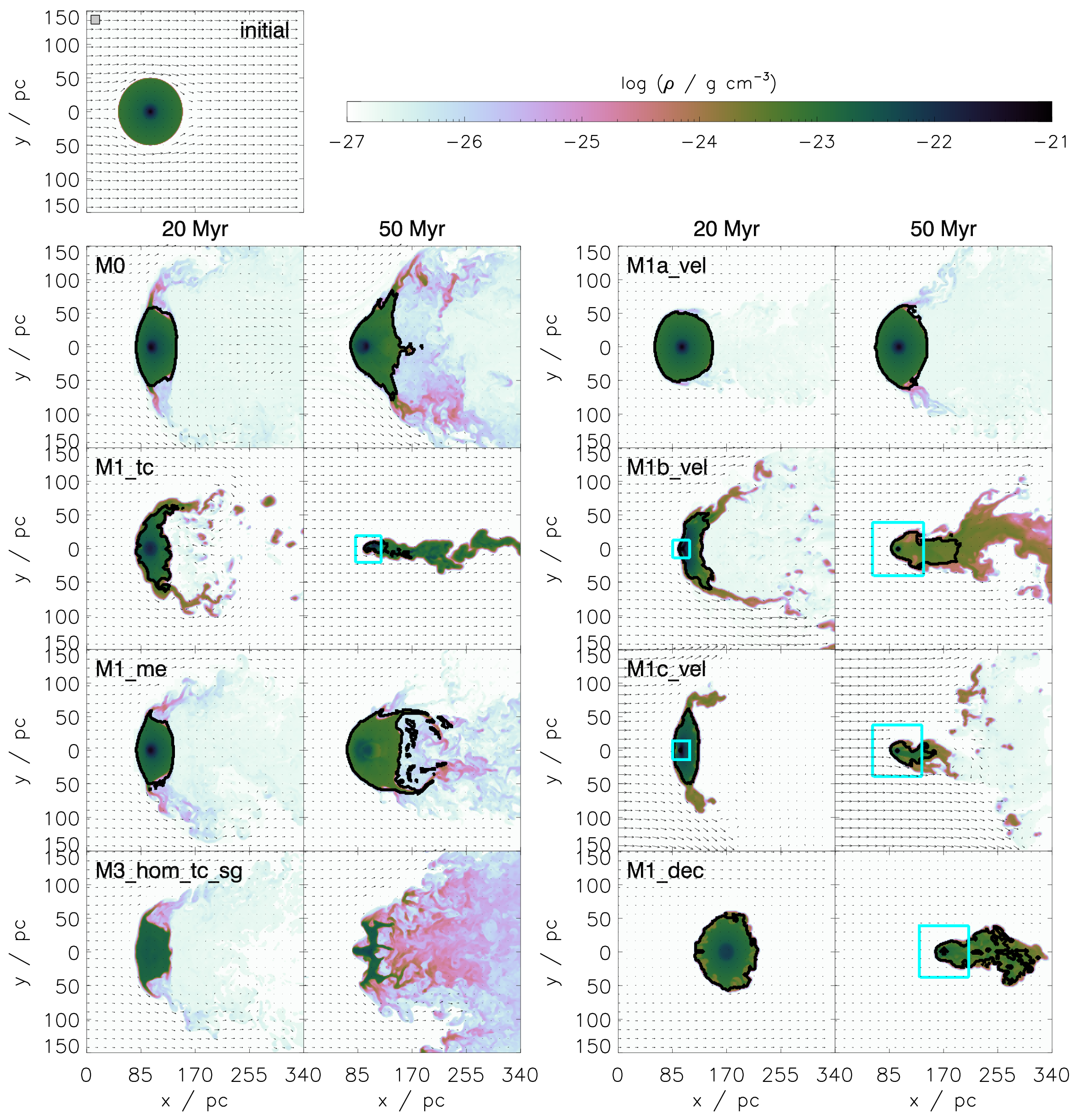}
\caption{Evolution of density in the massive model clouds. The plots show the section plane (x,y) through cloud centre. The \emph{upper most plot} shows the initial stage. The evolution is shown in snapshots after $20~$Myr (\emph{left column}) and after $50~$Myr (\emph{right column}). Model M3\_hom\_tc\_sg is shown at $55~$Myr, because the finger-like structures, which are due to \rti, are substantially developed. Model M1\_dec is shown at $100~$Myr when the head-tail structure is more prominently developed. The \emph{black solid lines} enclose cloud material (gravitationally bound and physically connected) and the \emph{light blue solid lines} border gravitationally instable regions. \emph{Black arrows} illustrate the motion of gas. The arrow lengths are scaled to $v_{\rm ini}=333\kms$ (model M1c\_vel). The \emph{grey square} in the upper left corner is spanned by $10\times 10$ cells of finest numerical resolution ($\Delta x=1.3~$pc).}
\label{fig:density-1}
\end{figure*}

\begin{figure*}
\centering
\includegraphics[width=\linewidth]{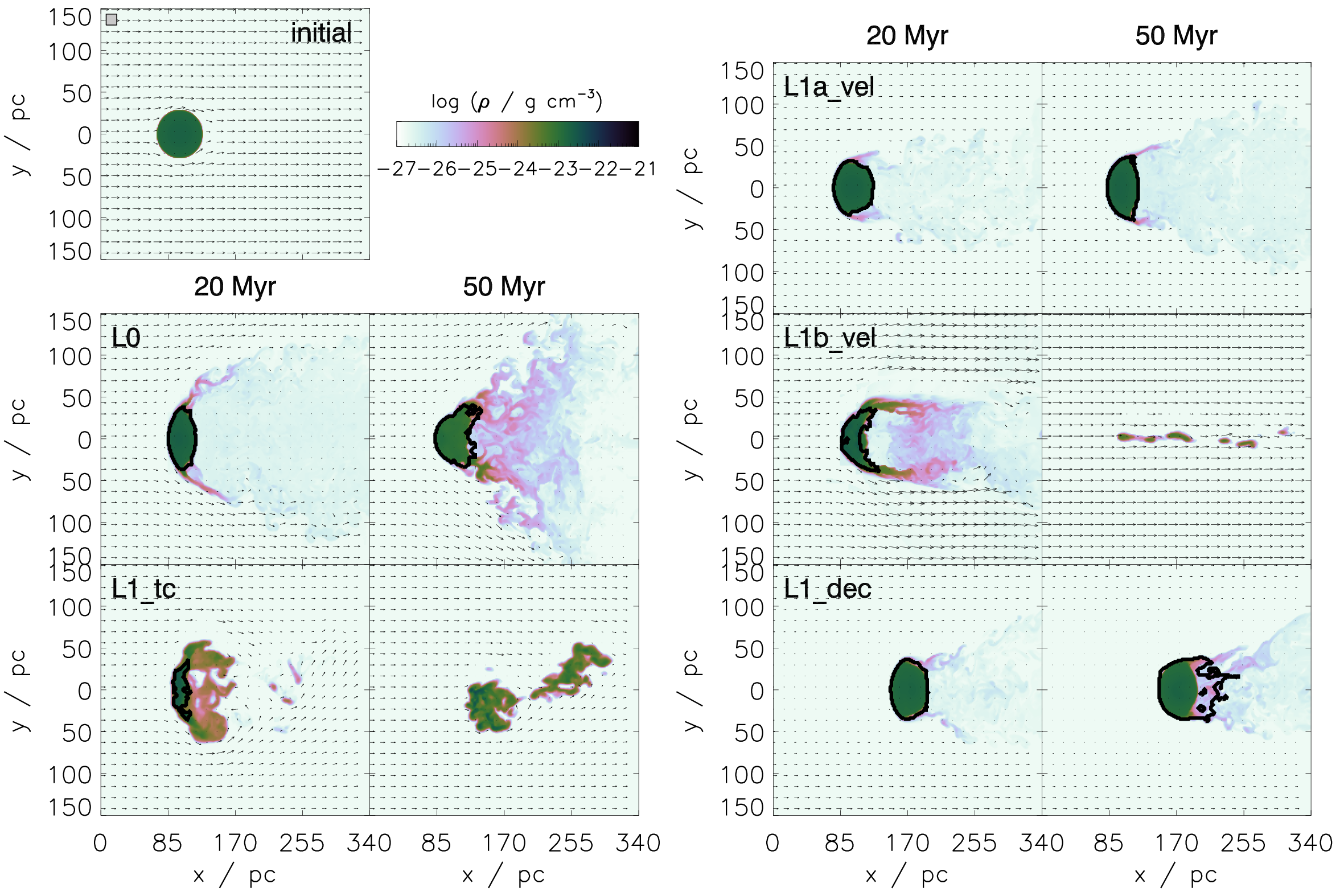}
\caption{The same as Fig.~\ref{fig:density-1}, but for the low-mass model clouds. The arrow lengths are scaled to $v_{\rm ini}=248\kms$ (model L1b\_vel).}
\label{fig:density-2}
\end{figure*}

In the middle plot of Fig.~\ref{fig:mass-1} it is striking that $\dot{M}>0$ at some times. By inspecting the respective plots for models M0 and L0 in Figs.~\ref{fig:density-1} and \ref{fig:density-2} one observes the velocity field behind the cloud to be turbulent and hence directed towards the cloud. So, already stripped cloud material is advected back to the cloud by the turbulent rear wake. This effect is rather low, such that on average both models lose mass. The stripping is continuous, i.e. without any sudden peaks in mass loss (Fig.~\ref{fig:mass-1}), filamentary, and spatially extended (Figs.~\ref{fig:density-1} and \ref{fig:density-2}).

\subsection{Thermal conduction}\label{subsec:conduction}
\begin{figure}
\centering
\includegraphics[width=\linewidth]{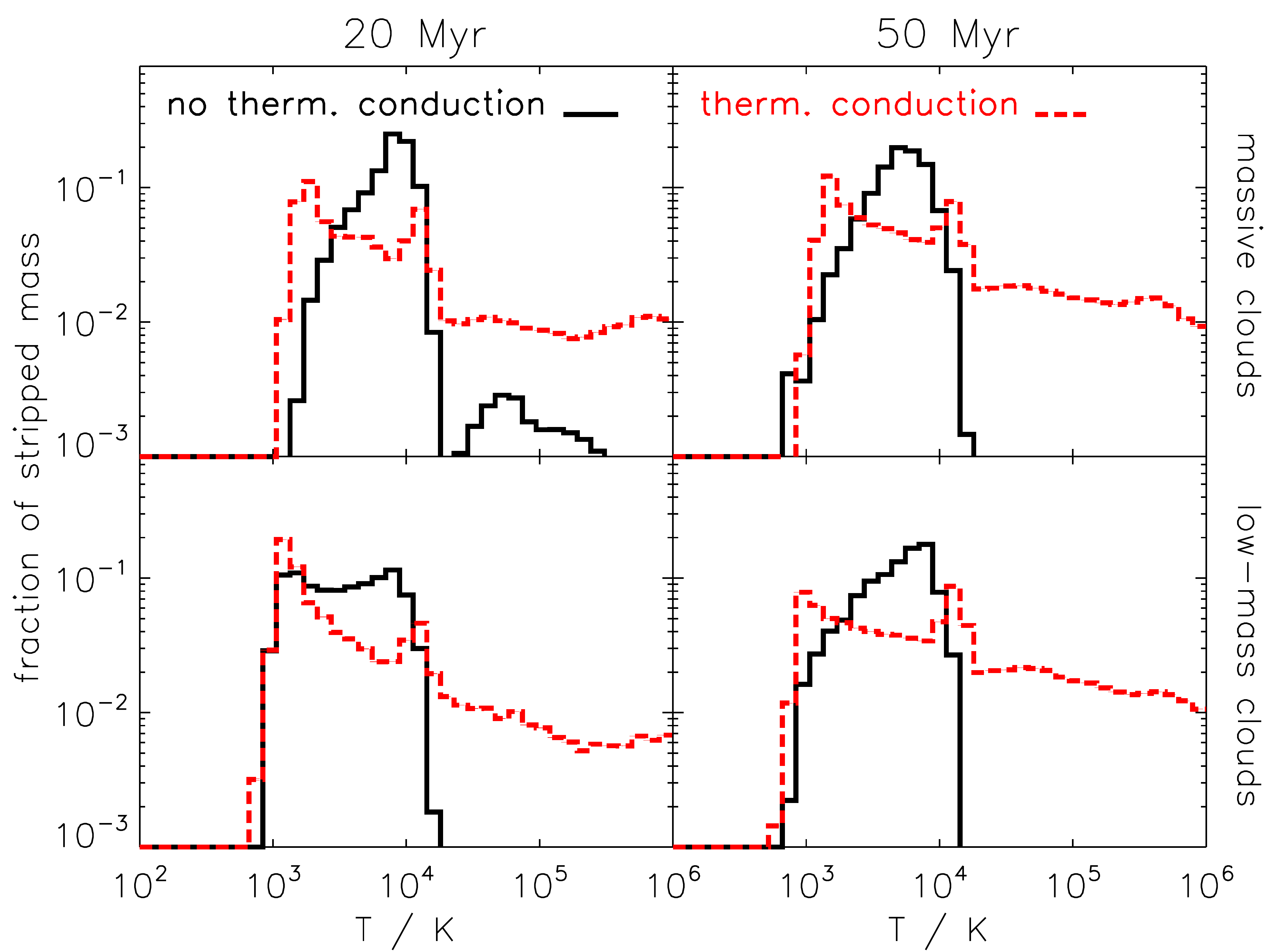}
\caption{Distributions of temperature in the stripped material of massive clouds (\emph{upper panel}) and low-mass clouds (\emph{lower panel}) with thermal conduction (\emph{red dashed lines}) and without thermal conduction (\emph{black solid lines}) after $20~$Myr (\emph{left column}) and after $50~$Myr (\emph{right column}). The standard deviation in every bin is too low to be visible in the plots.}
\label{fig:conduction-5}
\end{figure}

By comparing models M0 with M1\_tc (Fig.~\ref{fig:density-1}) and L0 with L1\_tc (Fig.~\ref{fig:density-2}) one easily verifies the different evolution of clouds that differ in thermal conduction only. Material is stripped from clouds without thermal conduction in form of dense clumps that are confined and isolated from each other. From Fig.~\ref{fig:conduction-5} we learn that these stripped clumps cool efficiently, because particle density is sufficiently high. The sharp cut around $2\times 10^4~$K correlates with a maximum of the cooling strength, because hydrogen gets fully ionized. Thus, the budget of thermal energy in stripped fragments is regulated by radiatve cooling. In contrast, the material stripped from clouds with thermal conduction is too dilute so it is heated up quickly above $10^5~$K, which conforms qualitatively with the findings of \citet[][]{17armillottaetal}. Irrespective of thermal conduction almost all of the stripped gas is above $10^3~$K, i.e. it is fully dissociated.

\begin{figure}
\centering
\includegraphics[width=\linewidth]{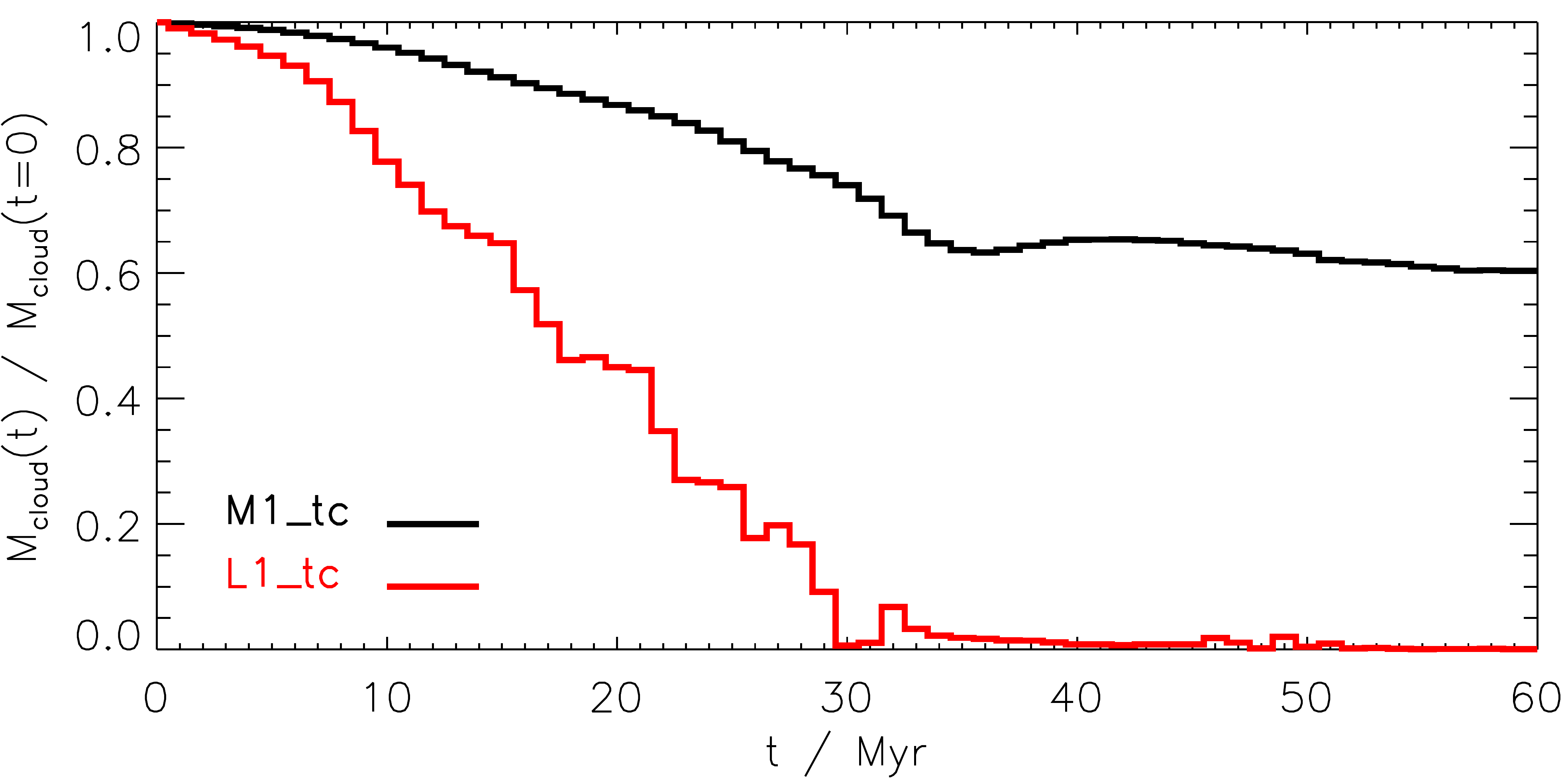}
\caption{The same as Fig.~\ref{fig:mass-1}, but for respective clouds without thermal conduction.}
\label{fig:conduction-6}
\end{figure}

Massive clouds without thermal conduction undergo a Jeans instability, while massive clouds with thermal conduction do not. In contrast, low-mass clouds do not experience any Jeans instability irrespective of thermal conduction (models M0, M1\_tc, L0, L1\_tc in Figs.~\ref{fig:density-1} and \ref{fig:density-2}). Low-mass clouds without thermal conduction are already disrupted after $30~$Myr (Fig.~\ref{fig:conduction-6}) while those with thermal conduction still retain more than $50~$per cent of their initial mass (Fig.~\ref{fig:mass-1}). The mean mass-loss rates of massive clouds without thermal conduction are $\sim 33~$per cent higher than in their counterparts with thermal conduction. In low-mass clouds this excess is even $\sim 50~$per cent (Table~\ref{table:massloss-2}), which agrees qualitatively with the results of \citet[][]{07vieserhensler1}, and implies a longer life-time for clouds with thermal conduction. Thermal conduction thus acts in two ways: it prevents massive clouds from becoming gravitationally instable and it prevents low-mass clouds from being tattered.

By comparing models M0 and M1\_tc in Fig.~\ref{fig:density-1} it is striking that the central region in the massive cloud without thermal conduction becomes gravitationally instable after $29~$Myr (Table~\ref{tab:grav-1}). Because the density and hence the mass increase significantly in the central region, the mass stripped by ram-pressure can be compensated (Fig.~\ref{fig:conduction-6}). Consequently, the difference in mass loss between models M0 and M1\_tc starts to level out after $\sim 35~$Myr. 

\subsection{Cloud velocity}\label{subsec:velocity}
In addition to the amount of initial mass and thermal conduction also the relative speed between C\hvc{} and \cgm{} strongly influences the evolution of clouds. By inspecting models M0, M1a\_vel, M1b\_vel, M1c\_vel (Fig.~\ref{fig:density-1}) and L0, L1a\_vel, L1b\_vel (Fig.~\ref{fig:density-2}) a completely different morphology is seen after $50~$Myr. The second column in Table~\ref{table:massloss-2} indicates that all clouds lose the more gas the faster they move. In particular, low-mass clouds moving at $165\kms$ (Mach $0.46$) or below survive the first $50~$Myr of evolution with $65~$per cent of their initial mass, but when moving at $248\kms$ (Mach $0.69$) or above they are totally disrupted before $40~$Myr. However, massive clouds above $250\kms$ (Mach $0.69$) start to locally undergo a gravitational collapse in their central regions.

These very fast model clouds develop a so-called \emph{head-tail morphology}, i.e. an asymmetric density distribution with a sharp gradient at the cloud head and a shallow gradient in the trailing part. The stripped trail of massive clouds is the more spatially confined the faster the clouds move (Fig.~\ref{fig:density-1}). Such head-tail structures are actually observed at many C\hvcs{} \citep[][]{11putmansaulmets,11winkeletal}. Two prime examples are \hvc{}~$125$+$41$-$207$ \citep[][]{01bruenskerppagels} and \hvc{}~$297$+$09$-$253$ \citep[][]{06benbekhtietal}, two C\hvcs{} that move at $v_{\rm LSR}=-207\kms$ and $v_{\rm LSR}=-253\kms$, respectively. Oppositely, the low-mass model clouds do not develop such a prominent tail.

\begin{figure}
\centering
\includegraphics[width=\linewidth]{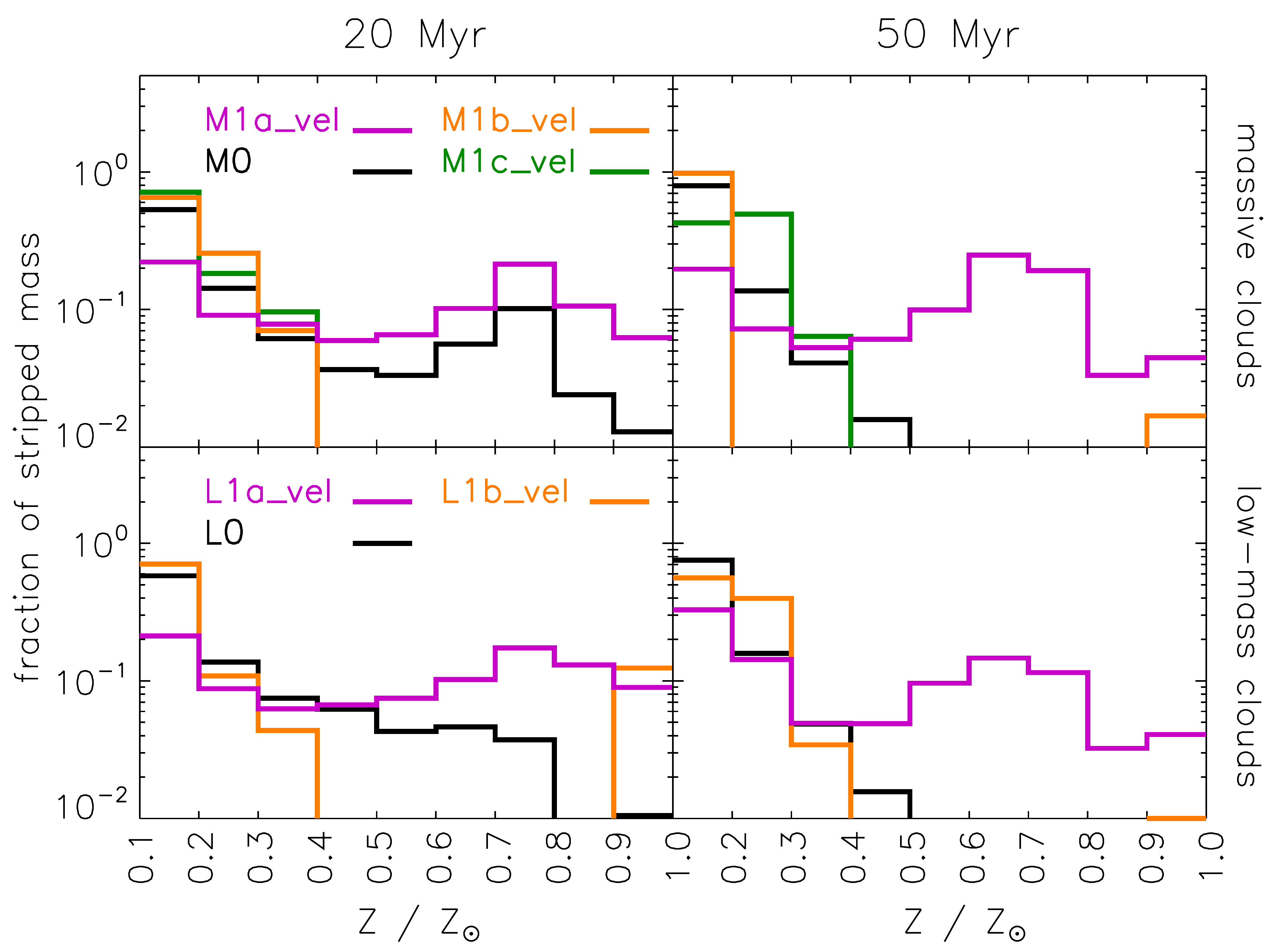}
\caption{Distributions of metallicity in the stripped material of massive clouds (\emph{upper panel}) and low-mass clouds (\emph{lower panel}) after $20~$Myr (\emph{left column}) and after $50~$Myr (\emph{right column}).}
\label{fig:velocity-5}
\end{figure}

The model clouds with a sub-solar metallicity of $0.1\zsolar$ can be used to study the impact of velocity on the spatial distribution of metals. From Fig.~\ref{fig:velocity-5} we learn that in fast clouds (L1b\_vel, M1b\_vel, M1c\_vel) the stripped material is mostly at initial cloud metallicity $0.1\zsolar$. Contrary, in slow clouds (L0, L1a\_vel, M0, M1a\_vel) stripped material has time to get enriched with \cgm{} metals. Hence, a higher fraction of stripped material is at higher metallicity. 

As can be seen in Fig.~\ref{fig:velocity-5} the overall metallicity in the stripped gas of models M0 and L0 is lowered at $50~$Myr compared to $20~$Myr. The reduced metallicity is due to higher mass-loss rates around $50~$Myr than at $20~$Myr (middle plot in Fig.~\ref{fig:mass-1}). Hence, more cloud material with $Z=0.1\zsolar$ streams into the trailing parts of the clouds. For M0 a positive linear correlation of $0.36$ between mass-loss rate and mean metallicity of the stripped gas is found at significance level of $0.05$. For L0 there is no such correlation between mass-loss rate and metallicity in the tail.

\begin{figure}
\centering
\includegraphics[width=\linewidth]{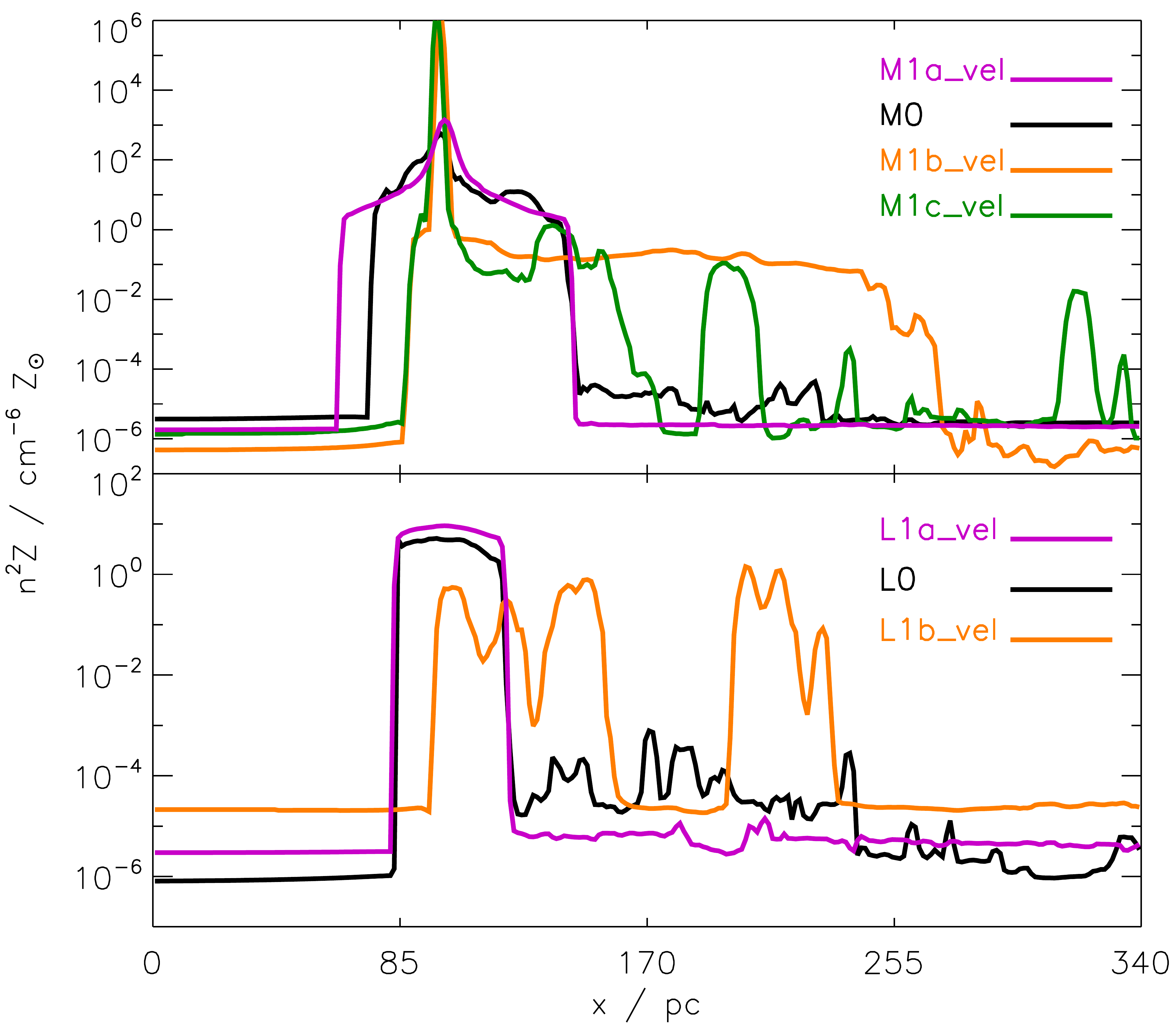}
\caption{Metallicity profiles weighted with the square of particle density along the direction of motion for the massive clouds (\emph{upper plot}) and low-mass clouds (\emph{lower plot}) after $50~$Myr.}
\label{fig:velocity-10}
\end{figure}

The mixing with \cgm{} appears dominantly at rear side of cloud. In Fig.~\ref{fig:velocity-10} the rear side of clouds are at higher metallicity than the front side. Also in Fig.~\ref{fig:velocity-10} we observe a lower metallicity behind high-speed clouds, because the region behind the cloud is refilled quickly with new cloud material.
By inspecting Figs.~\ref{fig:density-1} and \ref{fig:density-2} another morphologic feature becomes apparent in model clouds M1b\_vel and L1b\_vel, which is called a \emph{bow-shock shape}, i.e. the cloud is shaped like a boomerang with a density gradient along the wings. Model M1c\_vel develops this morphology later at $\sim 35~$Myr, but none of the slower clouds forms such a shape. Two observed examples with bow-shock shape are C\hvc{}~$172$-$60$ and C\hvc{}~$157$+$03$, which move at $v_{\rm LSR}=-235\kms$ and $v_{\rm LSR}=-184\kms$ \citep[][]{04westmeierbruenskerp,05westmeierbruenskerp}. Their shapes are interpreted as a consequence of ram-pressure by an ambient medium.

\begin{figure}
\centering
\includegraphics[width=\linewidth]{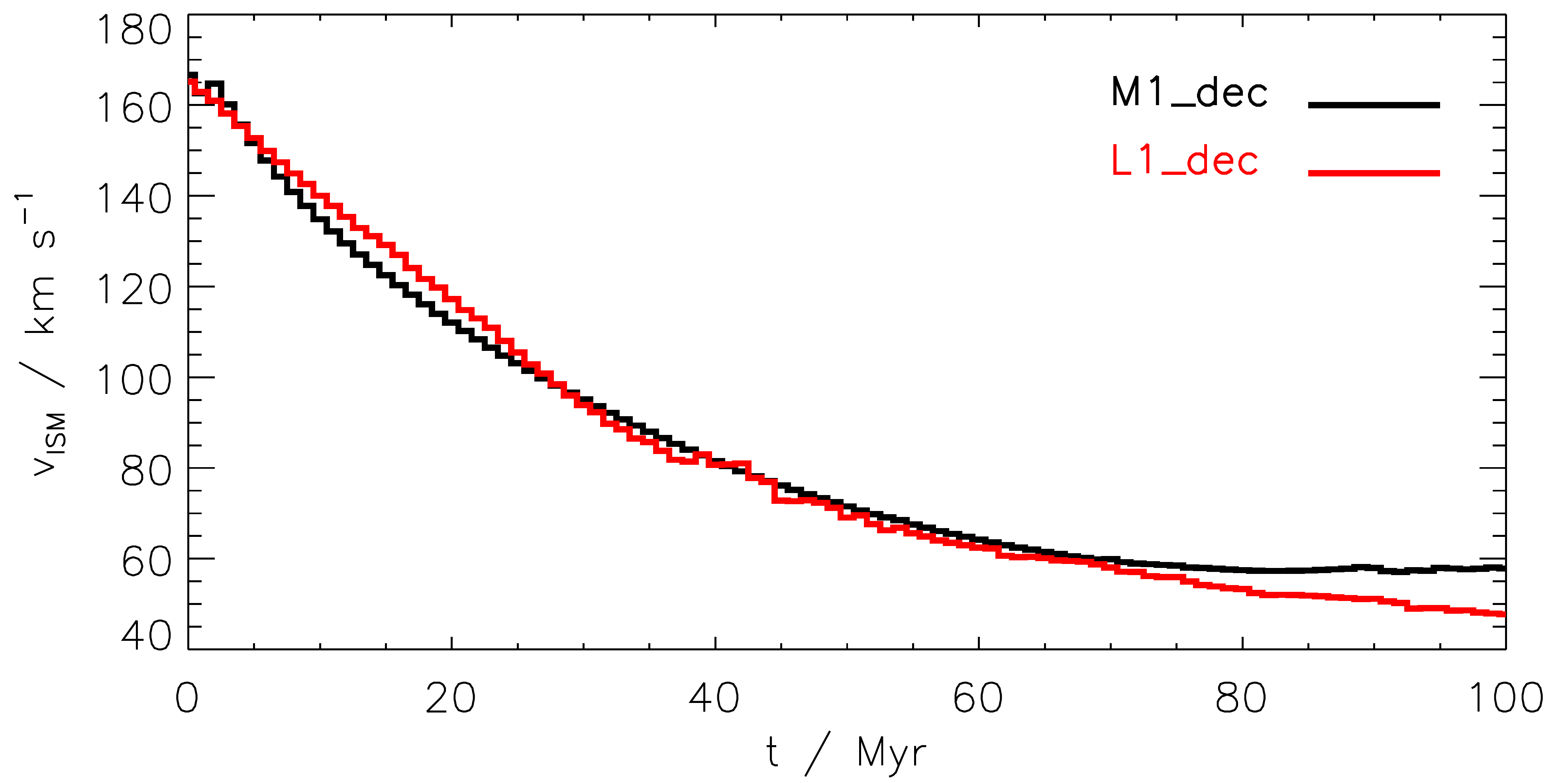}
\caption{Deceleration of model clouds that experience drag.}
\label{fig:velocity-9}
\end{figure}

With models M1\_dec and L1\_dec we are able to study the effect of drag from ambient gas, which leads to a deceleration of C\hvcs. In Figs.~\ref{fig:density-1} and \ref{fig:density-2} it is striking that decelerated, massive clouds can become gravitationally instable. They develop a head-tail morphology much later than the respective clouds that move at a constant speed. The mass loss is dramatically reduced, irrespective of initial mass (see Table~\ref{table:massloss-2}). In Fig.~\ref{fig:velocity-9} we show that the terminal velocity of M1\_dec is reached after $\sim 80~$Myr and reads $\sim~58\kms$. The terminal velocity of L1\_dec is still not reached after $100~$Myr, where the cloud is decelerated to $\sim 48\kms$. The mean deceleration rates are $-1.1\kms~$Myr$^{-1}$ and $-1.2\kms~$Myr$^{-1}$, respectively. So both models are decelerated well below the velocity threshold for \hvcs{} of $v_{\rm LSR}=100\kms$ thus showing a transition to intermediate-velocity clouds (\ivcs).

\subsection{Radiative cooling}\label{subsec:cooling}
\begin{figure}
\centering
\includegraphics[width=\linewidth]{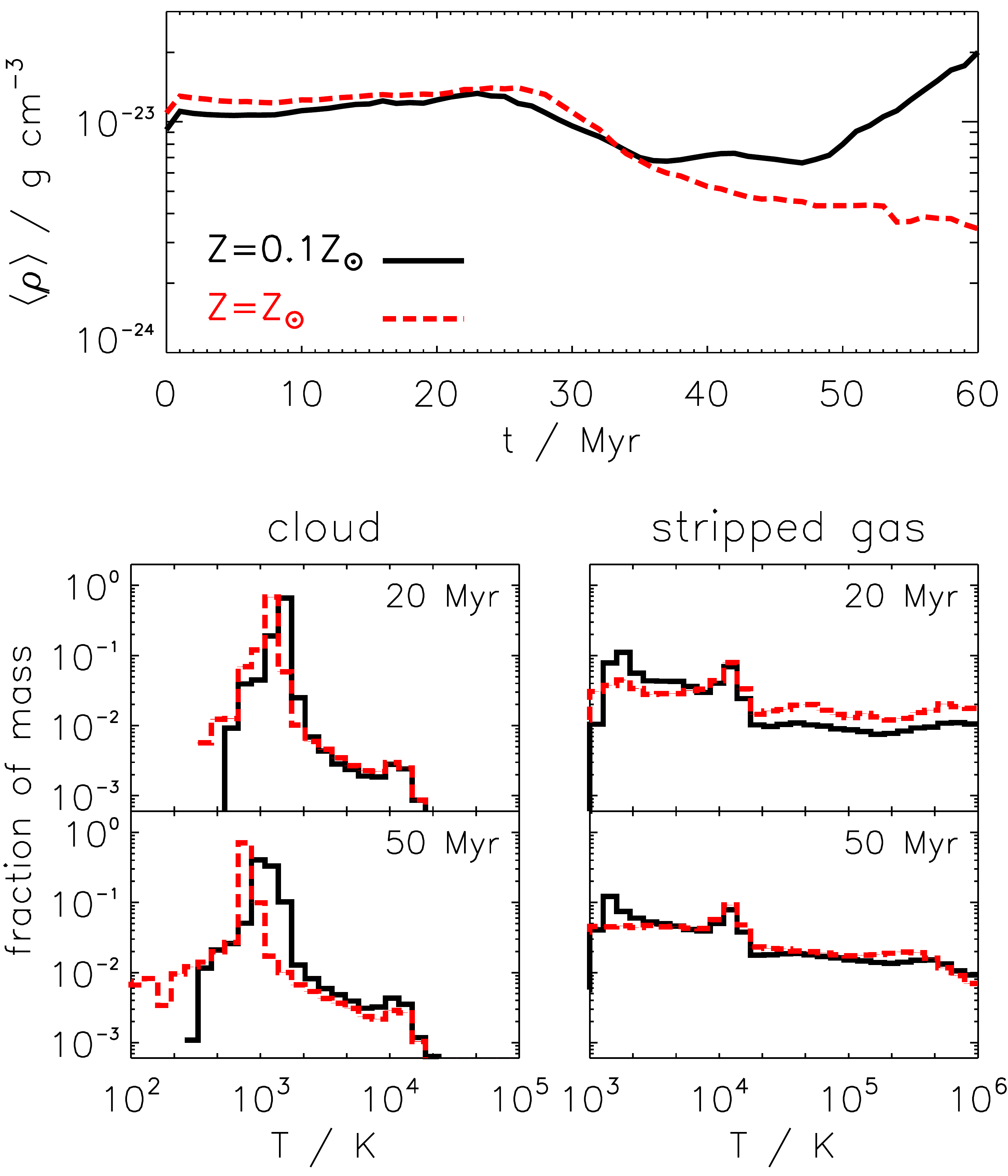}
\caption{\emph{Upper panel:} Mean density of model clouds M0 (\emph{black solid line}) and M1\_me (\emph{red dashed line}). \emph{Lower panel:} Distributions of temperature within the clouds (\emph{left column}) and in the stripped material (\emph{right column}) after $20~$Myr (\emph{upper row}) and after $50~$Myr (\emph{lower row}). The standard deviation in every bin is too low to be visible in the plots.}
\label{fig:cooling-2}
\end{figure}

We explore the effect of cooling strength on the evolution of C\hvcs{} by means of a different metallicity in the model clouds M0 ($0.1\zsolar$) and M1\_me ($\zsolar$). Because thermal conduction is active in both clouds, the stripping is filamentary and hence the stripped material bears the same temperature distribution irrespective of metallicity, hence irrespective of cooling strength (Fig.~\ref{fig:cooling-2}, right column in lower panel). However, the cloud is cooler on average for $Z=\zsolar$ at $50~$Myr, even though the mean density inside cloud M1\_me is significantly below that of M0 after $35~$Myr (upper panel in Fig.~\ref{fig:cooling-2}). The left column in the lower panel in Fig.~\ref{fig:cooling-2} shows that the peak of temperature distribution in M1\_me is below and for M0 it is above $10^3~$K. So, the metal-dependence of the cooling function is strong enough to have a measurable effect on cloud temperature.

\subsection{Physical simplicity}\label{subsec:simplicity}
Model M3\_hom\_tc\_sg is intentionally physically simplified. We neither consider thermal conduction, nor self-gravity and the cloud is homogeneous and isothermal. In the respective plot in Fig.~\ref{fig:density-1} it is observed that this cloud does neither develop a head-tail structure nor a bow-shock shape. It rather gets totally fragmented after $\sim 60~$Myr, because self-gravity does not stabilize the cloud against \rti{} thus the destructive modes can effectively grow. Such simplified cloud models do not maintain a density profile, which is a common feature being observed in C\hvcs.

So, simplified clouds evolve completely different than hydrostatic cloud models with self-gravity and thermal conduction. They are not realistic models in describing C\hvcs, because they do not develop the same features as observations reveal.

\section{Discussion}\label{sec:discussion}

\subsection{Column densities}\label{subsec:columndensities}
\begin{figure}
\centering
\includegraphics[width=.48\textwidth]{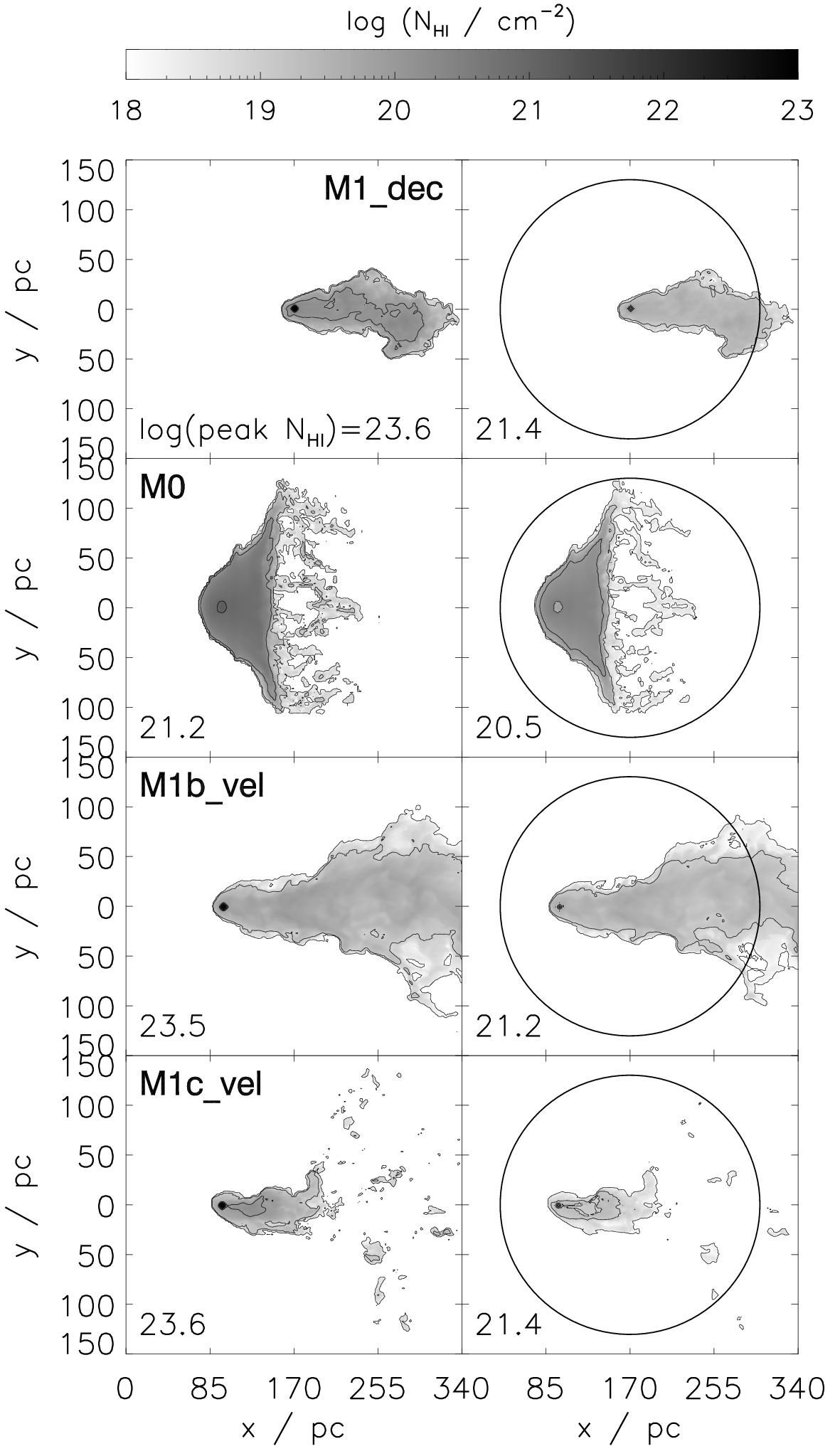}
\caption{Column densities of \ion{H}{i} for selected massive clouds after $50~$Myr (M1\_dec after $100~$Myr). The \emph{black isodensity contours} represent $\log\left(N_\ion{H}{i}/\scm\right)=18,19,\ldots$ \emph{Left column:} Actual $N_\ion{H}{i}$ for the clouds. \emph{Right column:} Diluted $N_\ion{H}{i}$ as being obtained at a distance of $100~$kpc by a beam of opening angle $9~$arcmin. The \emph{black circle} shows the area covered by this beam.}
\label{fig:columndensities-1}
\end{figure}

\begin{table}
\caption{Observed C\hvcs{} with high column densities.}
\label{table:columndensities-1}
\centering
\begin{threeparttable}
\begin{tabular}{l l l}  
\hline
Object & $\log\left(N_\ion{H}{i}/~\text{cm}^{-2}\right)$ & Reference \\
\hline
six C\hvcs & $\lesssim 21.0$ & \citet[][]{00braunburton} \\
C\hvc{} $125$+$41$-$207$ & $20.1\ldots 21.0$ & \citet[][]{01bruenskerppagels} \\
HVC $291$+$26$+$195$ & $\lesssim 20.1$ & \citet[][]{06benbekhtietal} \\
HVC $297$+$09.253$ & $\lesssim 20.5$ & \citet[][]{06benbekhtietal} \\
\hline
\end{tabular}
\end{threeparttable}
\end{table}
We have examined the column densities of neutral hydrogen, $N_\ion{H}{i}$, as they would appear at an observers distance including beam dilution when using single-dish telescopes, which have spatial resolutions of several arcmin (e.g. Effelsberg, Green Banks, Parkes, Dwingeloo). By comparing column densities of observed C\hvcs{} (Table \ref{table:columndensities-1}) with our simulated massive clouds we find reasonable agreement when accounting for beam dilution (Fig.~\ref{fig:columndensities-1}, right column). Exemplarily, we have chosen a beam with opening angle of $9~$arcmin, like Effelsberg telescope, at a distance of $100~$kpc, which is a typical distance for C\hvcs. The diluted column densities of our simulated clouds are much lower (by $\sim 2$ orders of magnitude) depending on the filling factor of the cloud in the beam. By this, the very high peak column densities (Fig.~\ref{fig:columndensities-1}, left column) are no longer visible. With larger beams (e.g. Dwingeloo with $30~$arcmin) or at greater distances (e.g. \citet[][]{00braunburton} find a distance bracket of $0.5~$to$~1~$Mpc for C\hvc{} $125$+$41$-$207$) our simulated $N_\ion{H}{i}$ would appear even lower. Thus, by considering realistic beam dilutions the column densities of our model clouds do not contradict to observed C\hvcs{} even though lying at the upper edge of consistency.

\subsection{Survival of clouds}\label{subsec:survival}
It has been demonstrated in \S~\ref{sec:modelcomparisons} that it strongly depends on the combination of initial cloud mass, cloud velocity, and thermal conduction if C\hvcs{} are being disintegrated or not during their passage through the \cgm. There are several dynamical processes, which can lead to mass loss of the clouds.

Several theoretical mass-loss rates have been determined to which we compare the mass loss obtained in our simulations. \citet{77cowiemckee} determine the rate of mass loss for a non-gravitating cloud by evaporation only (cooling effects are neglected) as
\begin{equation}
\dot{m}_{\rm CM}=3.25\times 10^{18}\ncgm\sqrt{\tcgm}\rcloud^2{\rm [pc]}\Phi F\left(\sigma_0\right).\label{equ:survival-1}
\end{equation}
The strength of saturation is given by
\begin{equation}
F\left(\sigma_0\right)=2\left[(\sigma_0H)^{1+Ma^2}\exp\left\{-2.5Ma^2\right\}\right]^{1/(6+Ma^2)},\label{equ:survival-2}
\end{equation}
where 
\begin{equation}
H=\left\{
\begin{array}{ll}
h^h/(h-1)^{h-1} &,Ma\leq 1 \\
11.5 &,Ma> 1
\end{array}
\right.\quad,\quad h=\frac{11+Ma^2}{1+Ma^2},\label{equ:survival-3}
\end{equation}
and $Ma$ is the Mach number of the flow around our clouds (cf. Table~\ref{table:Initial_Setup_2}). \citet[][]{93daltonbalbus} extended these investigations by introducing a continuous transition between classical and saturated heat flux. By simultaneously solving their equations (25) and (30) they obtain the ratio $\omega\left(\sigma_0\right)$ (cf. Fig.~\ref{fig:parameter-1}) between saturated and classical mass-loss rate. Hence,
\begin{equation}
\dot{m}_{\rm DB} = \omega\left(\sigma_0\right)\,\dot{m}_{\rm class}=\omega\left(\sigma_0\right)\,\frac{16\pi\mu_{\rm\cgm}\kappa_{\rm\cgm}\rcloud}{25k_{\rm B}}, \label{equ:survival-4}
\end{equation}
with $\dot{m}_{\rm class}$ given by \citet[][see equ.~(16) therein]{77cowiemckee} and our values for $\sigma_0$ are shown in Table~\ref{table:Initial_Setup_2}. \citet{86hartquistetal} found
\begin{equation}
\dot{m}_{\rm H} = \left[\varrho_{\rm\cgm}v_{\rm\cgm}\left(\bar c_{\rm cloud}\mcloud\right)^2\right]^{1/3}Ma^{4/3}\label{equ:survival-5}
\end{equation}
for the mass loss of a cloud in a subsonic wind without considering self-gravity. As we consider self-gravity in all model clouds (except in model M3\_hom\_tc\_sg) we also compare to the mass-loss rates analytically calculated by \citet{97arthurlizano},
\begin{equation}
\dot{m}_{\rm AL}\approx \frac{1}{24G}\frac{\bar c_{\rm cloud}^4}{c_{\rm\cgm}}Ma,\label{equ:survival-6}
\end{equation}
for self-gravitating clouds being subject to a subsonic wind. Here, $\bar c_{\rm cloud}$ and $c_{\rm CGM}$ denote the average sound speed inside the cloud and in the \cgm, respectively. The mass loss in Equ.~(\ref{equ:survival-6}) accounts for the ablation by hydrodynamical instabilities.

Each of the definitions (\ref{equ:survival-1}), (\ref{equ:survival-4}), (\ref{equ:survival-5}), and (\ref{equ:survival-6}) is based on only some of the physical processes we consider in our simulations. Thus, our mean mass-loss rates $\langle\dot m\rangle_{\rm sim}$ are either under- or overestimated (Table~\ref{table:massloss-2}). These analytical mass-loss rates are hence not sufficient to deduce a reliable mass loss from multiphysics simulations of C\hvcs.
\begin{table}
\caption{Mean mass-loss rates as obtained from the simulations (\emph{2nd column}) compared to several analytical mass-loss rates given by Equs.~(\ref{equ:survival-1}), (\ref{equ:survival-4}), (\ref{equ:survival-5}), and (\ref{equ:survival-6}). All rates are given in units of $10^{-4}\msolar~$/~yr. Some analytical rates cannot be applied to models without thermal conduction or without self-gravity.}           
\label{table:massloss-2}
\centering          
\begin{tabular}{l c c c c c}  
\hline
Model & $\langle\dot m\rangle_{\rm sim}$ & $\dot{m}_{\rm CM}$ & $\dot{m}_{\rm DB}$ & $\dot{m}_{\rm H}$ & $\dot{m}_{\rm AL}$ \\ 
\hline
M0 & $-3.6$ & $-6.2$ & $-7.7$ & $-4.8$ & $-2.7$ \\
M1\_tc & $-4.8$ & $-$ & $-$ & $-2.7$ & $-0.2$ \\
M1a\_vel & $-0.2$ & $-8.6$ & $-9.5$ & $-1.9$ & $-2.9$ \\
M1b\_vel & $-6.2$ & $-1.9$ & $-3.3$ & $-4.9$ & $-0.2$ \\
M1c\_vel & $-20.4$ & $-1.7$ & $-2.8$ & $-9.6$ & $-0.5$ \\
M1\_dec & $-0.2$ & $-7.0$ & $-8.2$ & $-0.02$ & $-0.06$ \\
M1\_me & $-0.7$ & $-6.0$ & $-6.7$ & $-4.1$ & $-5.6$ \\
M3\_hom\_tc\_sg & $-4.4$ & $-$ & $-$ & $-4.7$ & $-$ \\
L0 & $-1.6$ & $-4.0$ & $-4.4$ & $-1.7$ & $-1.4$ \\
L1\_tc & $-2.4$ & $-$ & $-$ & $-0.3$ & $-0.3$ \\
L1a\_vel & $-0.1$ & $-4.3$ & $-4.6$ & $-0.8$ & $-2.9$ \\
L1b\_vel & $-3.5$ & $-4.1$ & $-4.4$ & $-2.9$ & $-1.0$ \\
L1\_dec & $-1.9$ & $-5.0$ & $-4.9$ & $>-0.01$ & $-0.06$ \\
L1\_me & $-0.5$ & $-1.5$ & $-1.1$ & $-0.03$ & $-0.1$ \\
\hline                  
\end{tabular}
\end{table}
The clouds are subject to both \khi{} and \rti{} as a consequence of hydrodynamic interactions with the ambient gas. The \khi{} leads to the amplification of perturbations with wave numbers 
\begin{equation}
k>k_{\rm min}\approx\frac{g\rhocloud}{\varrho_{\rm\cgm}v_{\rm rel}^2},\;{\rm if}\;\rhocloud\gg\varrho_{\rm\cgm},\label{equ:survival-7}
\end{equation}
where $k=2\pi/\lambda$, with $\lambda$ being the wavelength of the perturbation, $g$ is the gravitational acceleration at the cloud's surface, $\rhocloud$ and $\varrho_{\rm\cgm}$ are the densities of C\hvc{} and \cgm, respectively, and $v_{\rm rel}$ denotes the relative velocity between them. The above relation~(\ref{equ:survival-7}) assumes discontinuities in both density and velocity between C\hvc{} and \cgm. Though this is valid initially (cf. \S~\ref{subsec:initialprofiles}), the dissipative effect of thermal conduction broadens the discontinuities over a finite domain (\emph{transition zone}) hence converting them into very steep gradients. It has been shown by \citet[][see their appendix B.2]{07vieserhensler1} that a finite region can be defined in $k$-space, which consists of all modes leading to an amplification of \khi{}. A cloud is stable against \khi{} for all modes outside of that instability strip, whose size inversely scales with the width of the transition zone. The thicker the transition zone, the thinner the instability strip. Thus, thermal conduction is able to suppress \khi{} in terms of decreasing the number of destructive modes. Our simulated model clouds M1\_tc and L1\_tc are subject to strong \khi{} that ablate big clumps of gas from the clouds. The disrupting \khi{} waves can propagate through the entire cloud hence leading to growing modes of instability everywhere in the cloud. Whereas the respective counterparts M0 and L0 are only mildly disturbed by \khi{} thus showing a continuous, filamentary, and low-density mass loss (respective plots in Figs.~\ref{fig:density-1} and \ref{fig:density-2}). One can deduce that thermal conduction extends the life-time of even weakly gravitationally bound clouds by lowering the mass loss. 

\rti{} can develop at the front side of the C\hvc{} when its initial density configuration is disturbed by the subsonically streaming \cgm, such that cloud gas is accelerated opposite to the direction of the density gradient in the cloud. If, like in model cloud M3\_hom\_tc\_sg, self-gravity does not stabilize the cloud surface against \rti{} \citep{93murrayetal} these instabilities grow by time and the disturbed outskirts of the C\hvc{} start to form typical finger-like structures (Fig.~\ref{fig:density-1}). These fingers of the dense cloud point into the dilute \cgm{} in the opposite direction of motion. The wavelength of the dominant mode is $\lambda=33~$pc (the separation between two adjacent ``fingers'' in Fig.~\ref{fig:density-1}) and the growth rate is given by \citep{61chandrasekhar}
\begin{equation}
\omega^2=\left(kv\right)^2=ak\frac{\rho_{\rm\cgm}-\rhocloud}{\rho_{\rm\cgm}+\rhocloud},\label{equ:survival-8}
\end{equation}
with $k=2\pi/\lambda$ being the wave number of the perturbing mode and $v$ is its phase velocity. By measuring $v\sim 2\kms$ the acceleration of the cloud surface is computed from Equ.~(\ref{equ:survival-8}) as $a\sim -3\times 10^{-9}~$cm~s$^{-2}$. According to \citet{91kull} the e-folding time, i.e. the growing time for a certain mode, reads $\tau_{\rm e}=(ka)^{-1/2}$. Hence the time, at which the spatial extension of the RT structures increases by a factor $e$, is $\tau_{\rm e}\sim 2.5~$Myr. The inverse denotes the growth rate $\tau^{-1}_{\rm e}\sim 0.4~$Myr$^{-1}$.
Another dynamical effect the clouds are subject to is the stripping of material by ram pressure. \citet{00moriburkert} determine the stripping radius for a galaxy of particular mass and moving at a certain speed. They balanced the thermal pressure of the cloud by the total pressure of the ambient medium, which in our case is
\begin{equation}
P_{\rm\cgm}=P_{\rm thermal}+\varrho_{\rm\cgm}v^2,
\label{equ:flow-3}
\end{equation}
with $P_{\rm thermal}$ given by Equ.~(\ref{equ:simulations-1}). The stripping radius for a specific cloud is the distance from cloud centre at which
\begin{equation}
P_{\rm\cgm}>P_{\rm cloud}(r)=\frac{k\varrho(r)T(r)}{\mu(r)},
\label{equ:flow-4}
\end{equation}
is true. Contrary to \citet{00moriburkert} we consider self-gravity thus the cloud temperature is calculated by Equ.~(\ref{equ:initial-2}) while the corresponding density is constrained by thermal equilibrium. Equ.~(\ref{equ:flow-4}) indicates disruption for all low-mass clouds, except model L1a\_vel. However, according to Table~\ref{tab:massloss-1} also the low-mass models L0 and L1\_dec are not disrupted.

\begin{figure}
\centering
\includegraphics[width=\linewidth]{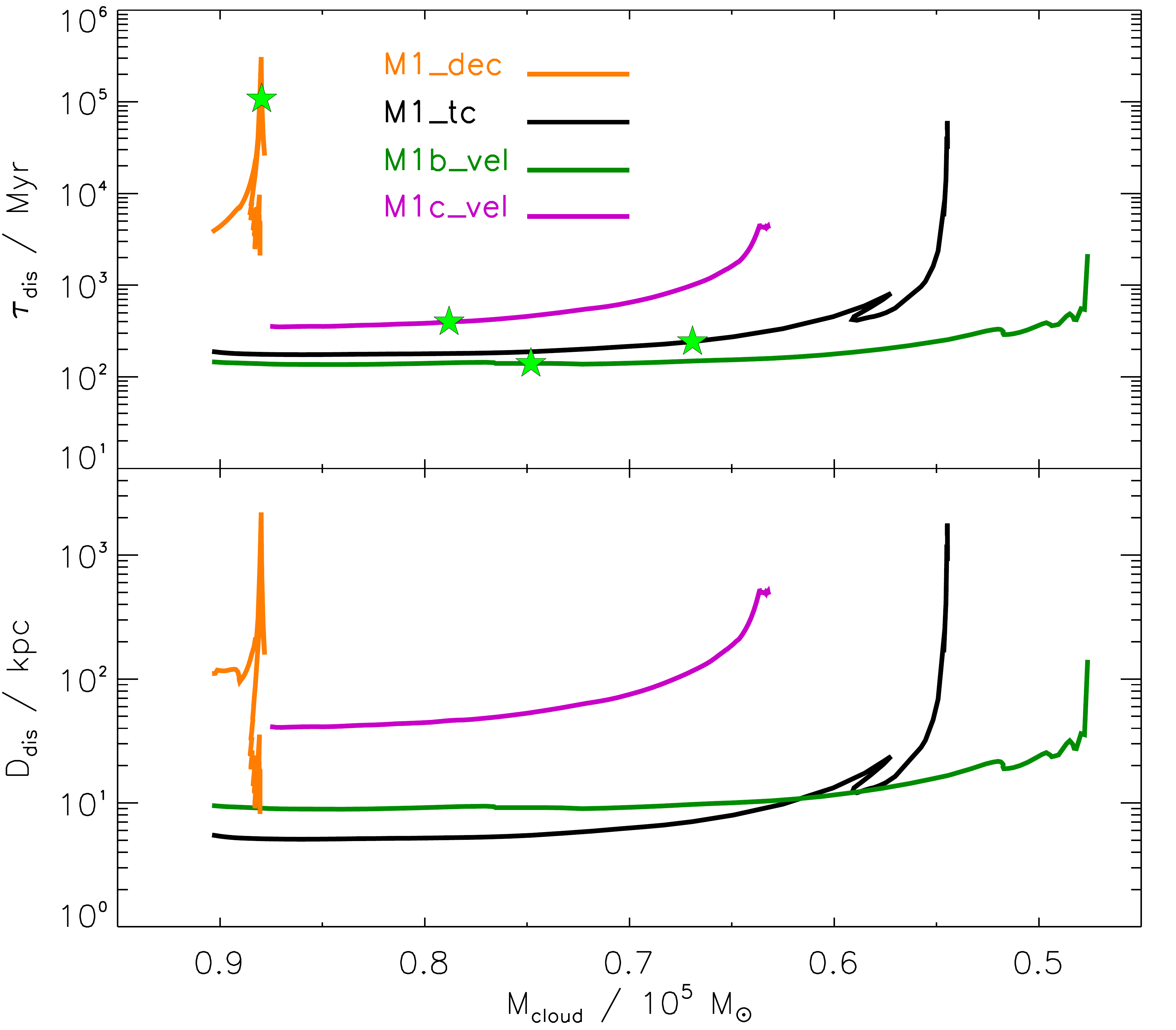}
\caption{Both disruption time $\tau_{\rm dis}$ and distance $D_{\rm dis}$ for clouds developing a gravitationally instable region. The \emph{light-green stars} indicate the onsets of collapse.}
\label{fig:massloss-1}
\end{figure}

Analogously to \citet{09heitschputman} we calculate the disruption time $\tau_{\rm dis}=\mcloud/({\rm d}M_{\rm cloud}/{\rm d}t)$ (time at which the cloud would be totally disrupted for a certain mass-loss rate) and disruption distance $D_{\rm dis}=\mcloud/({\rm d}M_{\rm cloud}/{\rm d}z)$ (travelled distance after which the cloud would be totally disrupted for a certain mass-loss rate) for all our clouds. In opposite to the clouds simulated by \citet[][]{09heitschputman} ours live much longer, because we assume much higher central densities and consider thermal conduction. Consequently, our mass-loss rates are much smaller than those in \citet[][]{09heitschputman} and so our values of $\tau_{\rm dis}$ and $D_{\rm dis}$ are greater than their values. In opposite to \citet[][]{09heitschputman} four of our clouds undergo a gravitational collapse (Table~\ref{tab:grav-1}). The values of $\tau_{\rm dis}$ and $D_{\rm dis}$ for these clouds increase over time (Fig.~\ref{fig:massloss-1}) indicating that these clouds will not be disrupted. Such a result has not been observed in the clouds simulated by \citet[][]{09heitschputman}. In \citet[][]{19sanderhensler} we have shown that self-gravity makes a distinct difference in the evolution of even homogeneous clouds with low densities and far below their Jeans masses.

Our massive clouds are comparable to model U in \citet[][]{07vieserhensler1}, who already found that the mass loss from a cloud with thermal conduction reduces after $60~$Myr of evolution (see \S~\ref{subsec:conduction}). This effect is even more pronounced in our simulations: for the entire simulation time the mass-loss rate of model M0 does not exceed $75~$per cent of the mass-loss rate of model M1\_tc (Fig.~\ref{fig:conduction-6} and Table~\ref{table:massloss-2}). After $47~$Myr the mass loss in model M1\_tc reduces, because the cloud centre collapses, which is accompanied by an increasing density.

\citet[][]{17armillottaetal} conducted simulations of C\hvcs{} moving through a hot \cgm. Their clouds with radii below $100~$pc do not survive, which can be a consequence of a rather weak gravitational potential, because their clouds with radius of $50~$pc contain only $160\msolar$, are homogeneous, isothermal, and do not consider self-gravity. Their clouds can most likely be compared to our model M3\_hom\_tc\_sg even though their initial cloud is obtained by adopting the cloud density accordingly to environmental conditions for obtaining pressure equilibrium. Hence, model M3\_hom\_tc\_sg is much more massive and compact as the clouds in \citet[][]{17armillottaetal}, but M3\_hom\_tc\_sg is already disrupted by \rti. Here, physical simplicity leads to non-realistic results.  

All of our model clouds, which are not disrupted after $50~$Myr, are compiled in Table~\ref{tab:massloss-1}.
\begin{table}
\centering
\caption{Model clouds that survive their passage through the \cgm{} for $50~$Myr. The 2nd column shows the remaining fraction of initial cloud mass after $50~$Myr.}
\begin{tabular}{l|c}
\hline
Model & $\mcloud(t=50)/\mcloud(t=0)$ \\
\hline
M0 & $0.85$ \\
M1\_tc & $0.63$ \\
M1a\_vel & $0.99$ \\
M1b\_vel & $0.56$ \\
M1c\_vel & $0.75$ \\
M1\_dec & $0.97$ \\
M1\_me & $0.93$ \\
L0 & $0.65$ \\
L1a\_vel & $0.97$ \\
L1\_dec & $0.98$ \\
\hline
\end{tabular}
\label{tab:massloss-1}
\end{table}

\subsection{Gravitational collapse}\label{subsec:grav}
In some of the C\hvcs, which resist all disrupting effects (Table~\ref{tab:massloss-1}), local Jeans instabilities can develop. Within our simulations all grid cells that belong to a cloud need to be identified. This is the case if a grid cell is

\begin{figure}
\centering
\includegraphics[width=\linewidth]{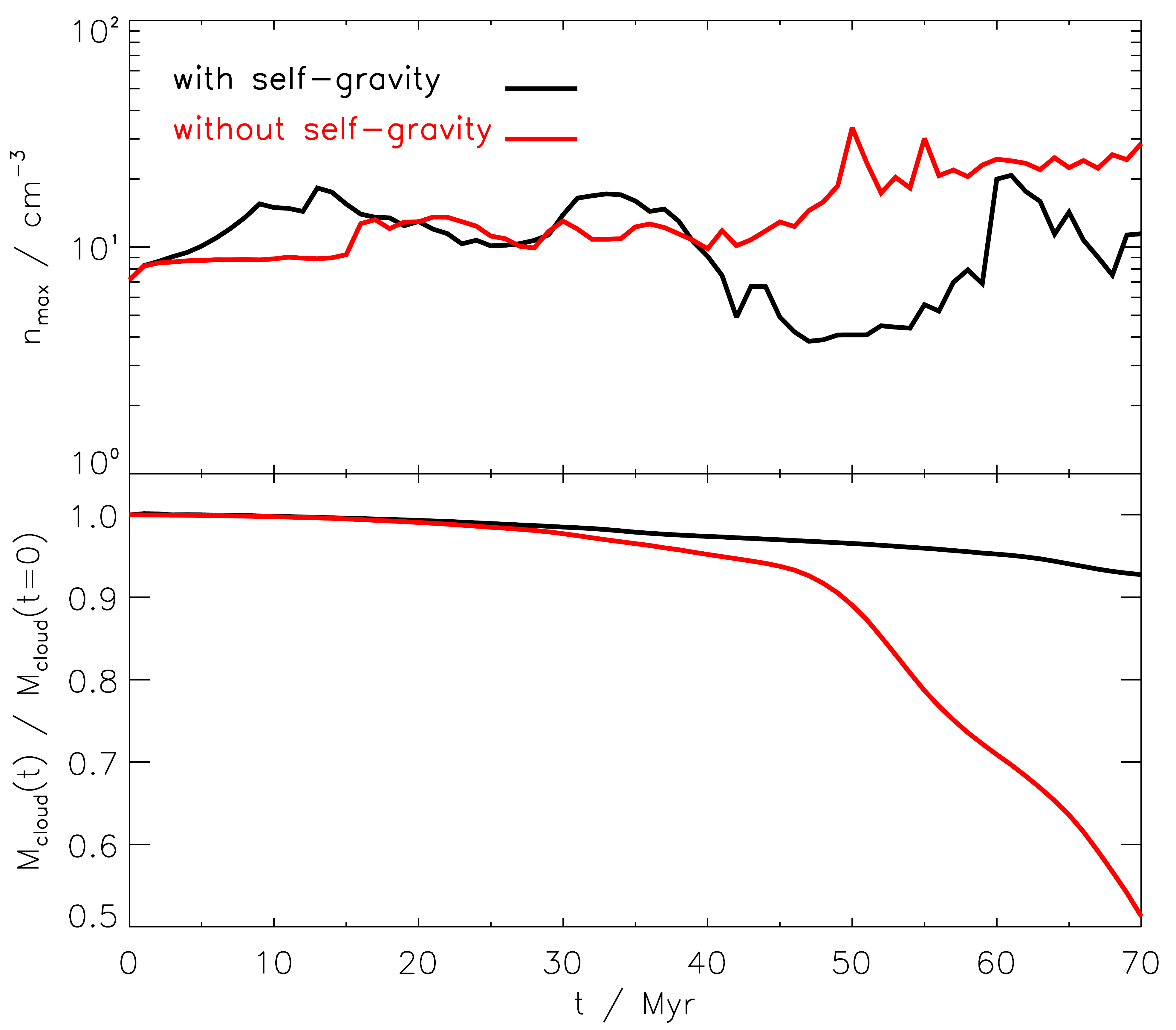}
\caption{Evolution of maximum particle density (\emph{upper plot}) and mass (\emph{lower plot}) of homogeneous and isothermal clouds with and without self-gravity.}
\label{fig:collapse-5}
\end{figure}

\begin{figure}
\centering
\includegraphics[width=\linewidth]{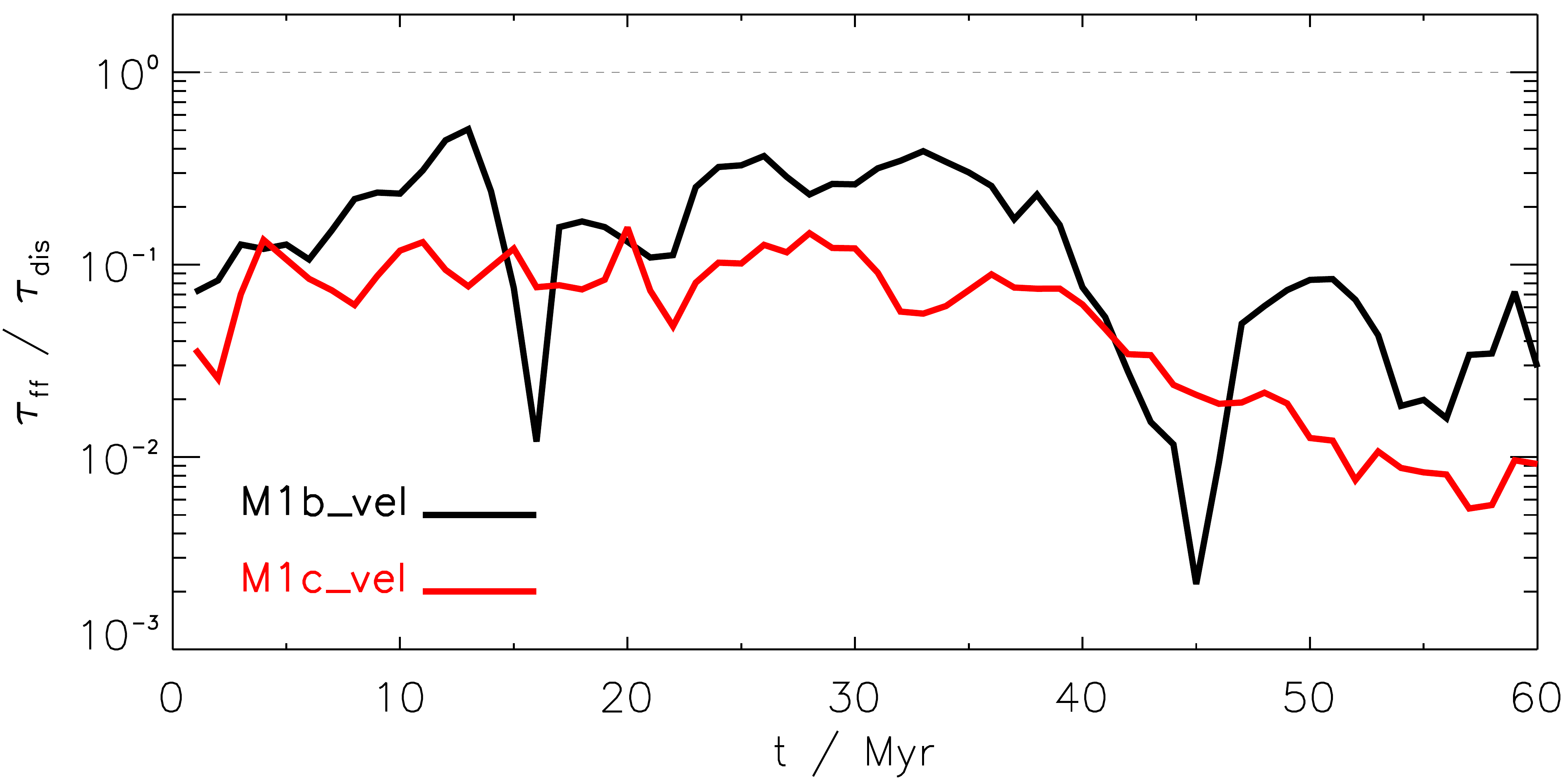}
\caption{Free-fall time versus disruption time in fastest massive clouds.}
\label{fig:collapse-4}
\end{figure}

\begin{enumerate}
\item gravitationally bound, i.e. the potential energy at the cell's location is greater than its kinetic energy, and
\item spatially connected to the cloud centre.
\end{enumerate}
In that regard, we do not account for grid cells, which are gravitationally bound, but are already stripped off from the cloud. The cloud centre is defined to be the cell with the global minimum in potential energy. In contrast, for model M3\_hom\_tc\_sg (no self-gravity) the above condition (i) is replaced by a threshold in density and temperature, i.e. $\rho_{\rm cell}>100~\rhocgm$ and $T_{\rm cell}<\tcgm$. To identify gravitationally instable regions we adapt the procedure described in \citet[][]{10federrathetal}:
\begin{enumerate}
\item Find all instable grid cells in a cloud by checking if $M_{\rm cell}>M_{\rm J}$. Take these cells as the centres of prospective collapsing regions, because these cells will have the potential minimum of this region.
\item Go radially outwards from each of the identified instable grid cells to their nearest neighbour cells and define this region to be the temporary control volume $VC$.
\item Check for instability of the control volume: $M_{\rm VC}>M_{\rm J}$.
\item Check for bounding of $VC$ by applying the virial theorem: $|E_{\rm grav}|\geq 2E_{\rm therm}$.
\item If checks (iii) to (iv) are successful, go back to step (ii) and repeat the procedure for the new $VC$. If one of the checks (iii) to (iv) fails, the current extension of the control volume defines the size of the gravitationally instable cloud region.
\end{enumerate}
By applying the above procedure to all surviving clouds (Table~\ref{tab:massloss-1}) we figured out four clouds (Table~\ref{tab:grav-1}), whose central regions become gravitationally instable despite the fact that all models are initially below their Bonnor-Ebert mass (Fig.~\ref{fig:initial-2}). \citet[][]{20lietal} performed a variety of simulations of cool clouds in the \cgm. For the sake of simplicity, their clouds are initially homogeneous and isothermal ($T\sim 10^4~$K) and are tuned to be in pressure balance with the ambient \cgm. They find that self-gravity is only relevant for Jeans-instable clouds. However, this conclusion is drawn from too short evolution times of $\leq 1~$Myr (see their figure 2). We do the same comparison for our simplified model M3\_hom\_tc\_sg by calculating it again with self-gravity. For a very early evolution ($\lesssim 2~$Myr) we find negligible differences between the clouds in both maximum particle density and mass loss (Fig.~\ref{fig:collapse-5}), which supports the observations of \citet[][]{20lietal}. However, for later stages ($>40~$Myr) differences become striking. Furthermore, in \citet[][]{19sanderhensler} we have shown that self-gravity leads to a core-halo density profile even for resting (i.e. without dynamical interaction with the ambient medium) and initially homogeneous clouds with mass below their Bonnor-Ebert mass. In opposite to \citet[][]{20lietal} the Jeans instabilities in our model clouds (Table~\ref{tab:grav-1}) occur for a broad range of cloud velocities and with or without thermal conduction. Hence we do not find any evidence that supports the reasoning by \citet[][]{20lietal} and we argue that self-gravity substantially affects the evolution of clouds with mass below their Bonnor-Ebert mass.
If massive clouds are fast enough (M1b\_vel, M1c\_vel), neither thermal conduction nor efficient mass loss can stop gravitational collapse. For both models the respective disruption time (\S~\ref{subsec:survival}) is always greater than the free-fall time (Fig.~\ref{fig:collapse-4}). We conclude that clouds with thermal conduction are able to collapse if they are very fast and sufficiently massive.

\begin{figure}
\centering
\includegraphics[width=\linewidth]{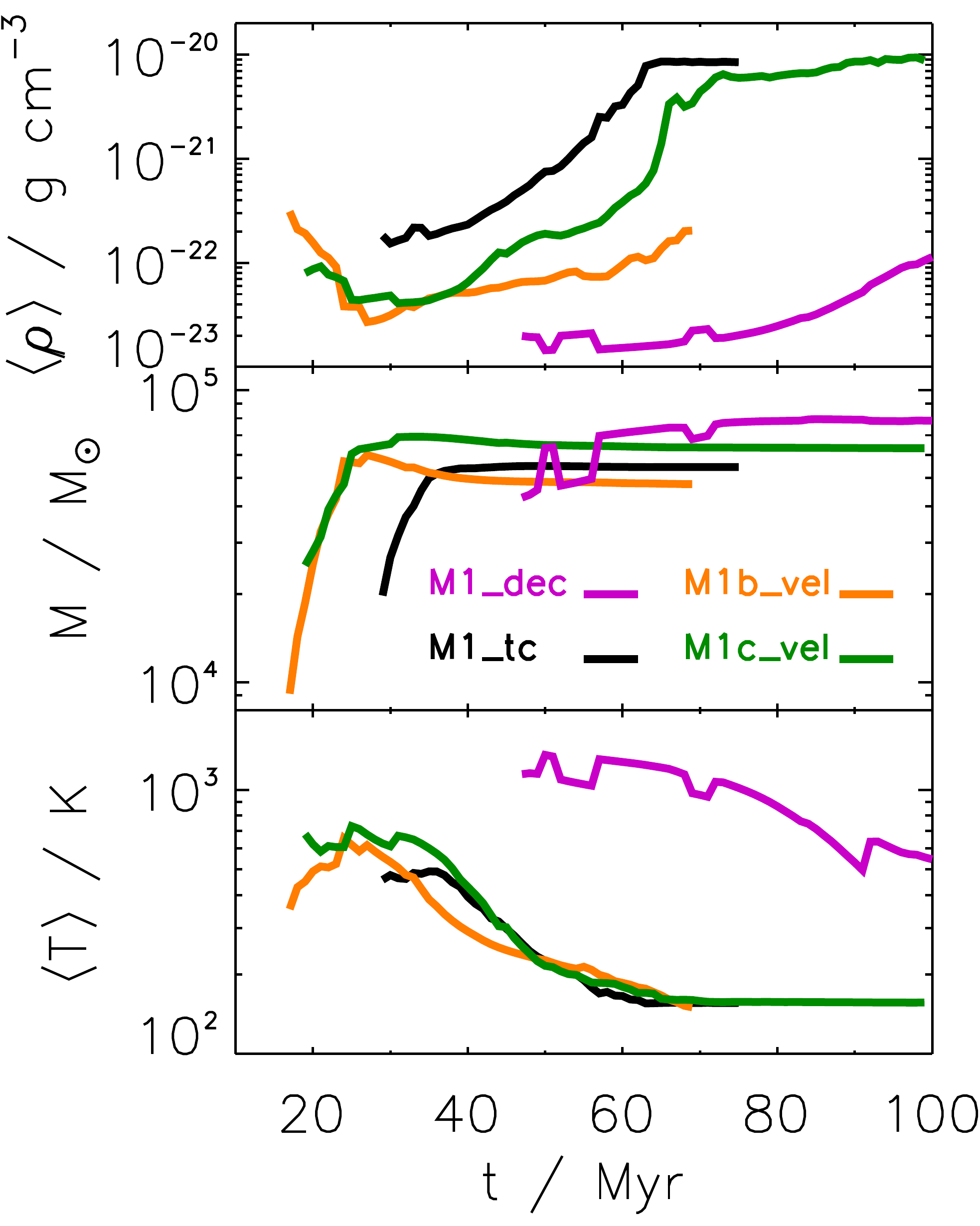}
\caption{Evolution of mean density (\emph{upper plot}), mass (\emph{middle plot}), and mean temperature (\emph{lower plot}) of gravitationally instable regions.}
\label{fig:collapse-1}
\end{figure}

The Jeans-instable regions in the four models evolve differently, which we discuss in the following by means of Fig.~\ref{fig:collapse-1} and Table~\ref{tab:grav-1}. The instable region in the most realistic model, M1\_dec, forms much later ($47~$Myr) than in the other three models, which is most likely due to both a decreasing dynamic pressure by deceleration and an enhanced thermal pressure in the cloud by thermal conduction. At this time, the cloud velocity has decreased below the threshold for \hvcs{} (Fig.~\ref{fig:velocity-9}).

We observe that the size of the instable region increases to roughly half of entire cloud size, but contains nearly all of cloud mass. The mean density is rather low and results in a gas surface-density below $0.01~$g~cm$^{-2}$. The thermal energy conducted into the instable region enhances the thermal pressure. At a certain point in evolution this cannot be compensated for by the decreasing dynamic pressure, which is due to cloud deceleration. The instable region thus expands. The density hence remains at low level and so is the cooling strength. Finally, the temperature in the instable region of model M1\_dec is high compared to those in the other model clouds. As a consequence, its mass does not substantially exceed the Jeans mass.

In the fast models with thermal conduction, M1b\_vel and M1c\_vel, the ram-pressure is constantly high during the entire evolution and so the instable regions are compressed to much higher densities. This situation is more pronounced for the faster cloud M1c\_vel, which is subject to a higher ram-pressure than model M1b\_vel. The temperatures reach values $\sim 150~$K thus lowering the Jeans mass of the instable region. Nonetheless, gas surface-densities are comparably low, too.

Model M1\_tc is distinct from the other three models. The entire cloud becomes gravitationally instable. Without thermal conduction the thermal pressure in the cloud is low such that at a lower speed as in models M1b\_vel and M1c\_vel the ram-pressure is sufficient to compress the cloud such that the density can reach $10^{-20}~$g~cm$^{-3}$. Being at a mean temperature of $\sim 150~$K the instable region greatly exceeds its Jeans mass. The gas surface-density reaches a value of $0.16~$g~cm$^{-2}$. According to \citet[][and references therein]{20traficanteetal} the threshold of gas surface-density for a region to form high-mass stars is $0.05$ to $0.1~$g~cm$^{-2}$ \citep[stellar feedback would increase this threshold to $1~$g~cm$^{-2}$,][]{08krumholzmckee}. Thus, our cloud M1\_tc would virtually be able to form even high-mass stars.

As has been discussed above, clouds with thermal conduction develop instable regions, but these do not get compact enough to become a star-formation site. For our studied cloud setups the instable regions are instead in some balance between gravity and thermal pressure. We conclude that thermal conduction might be a key process in preventing C\hvcs{} to form stars. These findings are in line with non-detections of stellar objects assigned to \hvcs{} in general in all hitherto performed surveys \citep[e.g.][]{01burtonbraunchengalur,03hoppschulteladbeckkerp2,03hoppschulteladbeckkerp,05siegeletal,07hoppschulteladbeckkerp}.

\begin{table}
\centering
\caption{Model clouds that experience a gravitational collapse with an onset at $t_{\rm collapse}$ (\emph{row 1}). The following rows tabulate parameters specific to the instable regions after respective $t_{\rm collapse}+40~$Myr: mass in terms of cloud mass (\emph{row 2}); size in terms of cloud size (\emph{row 3}); mass in terms of Jeans mass (\emph{row 4}); gas surface-densities (\emph{row 5}).}
\begin{tabular}{l c c c c}
\hline
& M1\_tc & M1b\_vel & M1c\_vel & M1\_dec \\
\hline
$t_{\rm collapse}$ / Myr & $29$ & $17$ & $19$ & $47$ \\
$M$ / $\mcloud$ & $1.00$ & $0.98$ & $1.00$ & $0.90$ \\
$V$ / $V_{\rm cloud}$ & $1.00$ & $0.58$ & $0.64$ & $0.51$ \\
$M$ / $\mjeans$ & $140.4$ & $8.0$ & $25.7$ & $1.6$ \\
$\Sigma$ / g~cm$^{-2}$ & $0.16$ & $<0.01$ & $0.02$ & $<0.01$ \\
\hline
\end{tabular}
\label{tab:grav-1}
\end{table}

\subsection{H$\alpha$ emission}\label{subsec:halpha}
A still puzzling feature of C\hvcs{} is their partially bright emission in the H$\alpha$ disexcitation-line. In the following we analyze how thermal conduction is able to contribute to the intensity of H$\alpha$ emission. We estimate the H$\alpha$ intensity emitted isotropically over $4\pi$ by any grid cell $j$ of our model clouds by
\begin{equation}
I_{\rm H\alpha}^{(j)}=A^{3\to 2}S^{3\to 2}_{\rm H\alpha}\;{\rm erg\;cm^{-2}\;s^{-1}},\label{equ:halpha-3}
\end{equation}
with $A^{3\to 2}=4.39\times 10^7~$s$^{-1}$ denoting the Einstein coefficient for spontaneous emission from $n=3\to 2$. By assuming ionization equilibrium (cf. \S~\ref{subsec:dission} and appendix \ref{app:dission}) the source function $S^{3\to 2}_{\rm H\alpha}$ is given by Boltzmann's Equation for excitation temperature. As the clouds are assumed to be optically thin, each of the $j$ contributions in Equ.~(\ref{equ:halpha-3}) has to be summed up to the total amount of $I_{\rm H\alpha}$. Thus, for $N$ cloud cells
\begin{equation}
I_{\rm H\alpha}^{\rm cloud}=\sum\limits_{j=1}^N I_{\rm H\alpha}^{(j)}\label{equ:halpha-4}
\end{equation}
yields the absolute value of H$\alpha$ intensity emitted by the cloud, which must be reduced by a dilution factor $\propto D^{-2}$ and corrected for the opening angle, $\Omega$, of the cloud when it is observed from a certain distance $D$ under a particular angle $\Omega$. A correction for an optically thick cloud with column density above $10^{18}$ cm$^{-2}$ is found in \citet[][cf. their equ.~(7)]{12bargeretal}. Furthermore, the emitted H$\alpha$ radiation must be corrected for extinction in the line-of-sight, which is highly uncertain \citep[][]{13hilletal}.

\begin{figure}
\centering
\includegraphics[width=\linewidth]{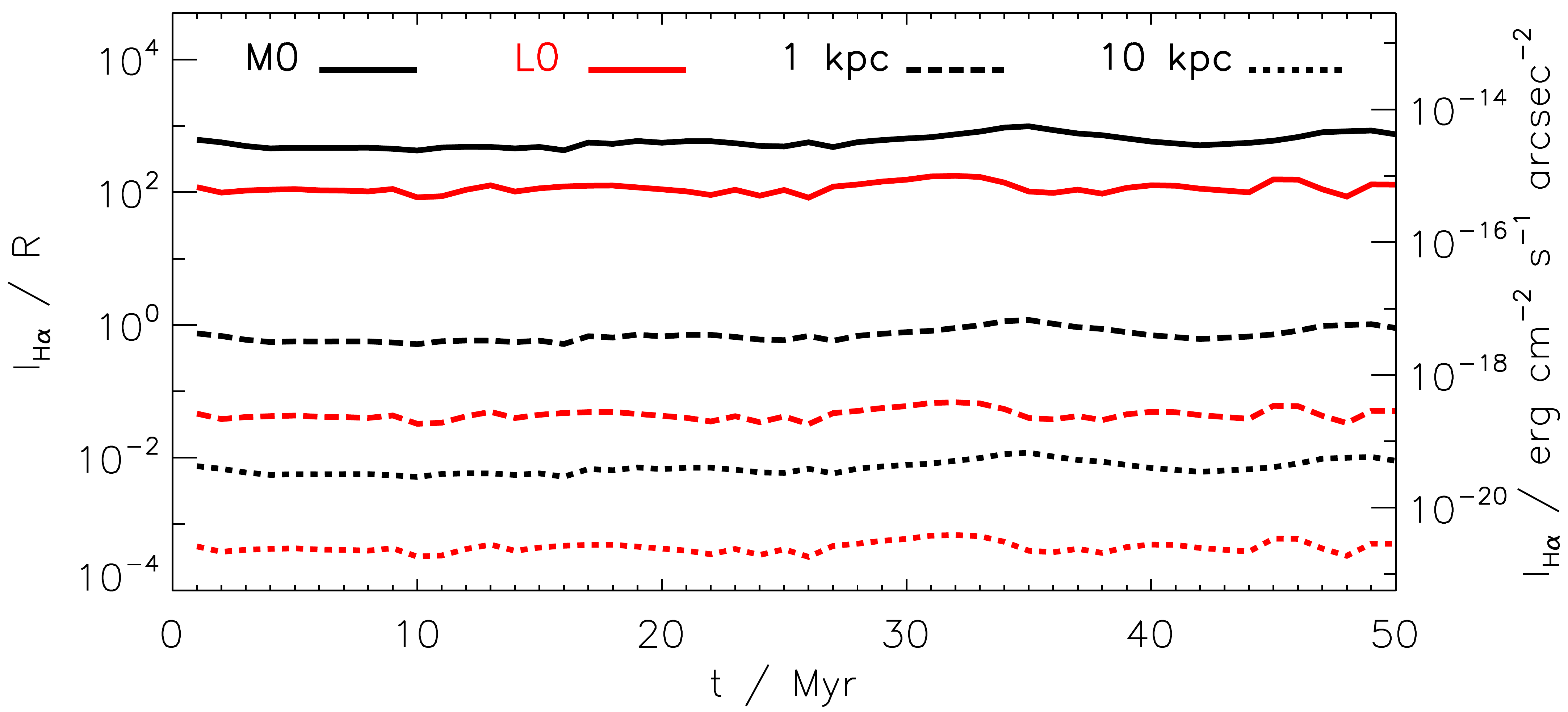}
\caption{Evolution of H$\alpha$ intensity as measured for the simulated reference clouds M0 (\emph{black solid line}) and L0 (\emph{red solid line}). The \emph{dashed-lines} and \emph{dotted lines} show the respective intensities at distances of $1~$kpc and $10~$kpc, respectively, in order to compare to \citet[][]{03putmanetal}.}
\label{fig:halpha-1}
\end{figure}

Observers mostly provide H$\alpha$ intensities in units of Rayleigh \citep[$1$R~$=2.41\times 10^{-7}~$erg~cm$^{-2}$~s$^{-1}$~sr$^{-1}$,][]{04vanwoerdenetal} or in units of erg~cm$^{-2}$~s$^{-1}$~arcsec$^{-2}$ (where sr$^{-1}=2.35\times 10^{-11}~$arcsec$^{-2}$). In Fig.~\ref{fig:halpha-1} the evolution of H$\alpha$ intensity~(\ref{equ:halpha-4}) is shown for the reference clouds M0 and L0 directly at cloud surfaces and at distances of $1~$kpc and $10~$kpc. For that we only consider dilution by distance, but no extinction in the line-of-sight. It is striking that the intensity of massive clouds is always higher than that of low-mass clouds, which is due to a greater number of contributing cloud cells. Interestingly, the H$\alpha$ intensity is nearly constant for both clouds. This can be understood in terms of two counteracting processes: the cloud loses mass thus the number of grid cells contributing to the intensity decreases. But simultaneously the cloud is heated up and so the intensity increases. Both effects roughly cancel out. 

When comparing to observed emission measures of H$\alpha$ from C\hvcs{} it turns out that the calculated intensities for our simulated clouds are at the same orders of magnitude for assumed distances of $>1~$kpc. \citet[][]{89kutyrevreynolds} report on H$\alpha$ emission from a \hvc{} in constellation Cetus with a best-fit mean value of $(8.1\pm 1.9)\times 10^{-2}~$R, which consists of two fields with $(7.3\pm 2.5)\times 10^{-2}~$R for field~1 and $(8.9\pm 2.8)\times 10^{-2}~$R for field~2. They infer the speed of the \hvc{} ($v_{\rm LSR} =-310\kms$) to be supersonic and the ambient gas density to be $n\lesssim (4-6)\times 10^{-3}\ccm$ in order to produce the observed shock-excited optical emission in H$\alpha$. Both the cloud speed and \cgm{} density in models M0 and L0 are below these values. \citet[][]{03putmanetal} performed deep H$\alpha$ spectroscopy towards several \hvcs, among which there are five C\hvcs{} with $20$ to $220~$mR at distances between $1.2$ and $13.2~$kpc. These clouds are exposed to a non-negligible ionizing UV background radiation in the Galactic halo due to $6~$per cent of UV photons that escape normal from the Galactic disk \citep[][]{02blandhawthornmaloney}. The calculated H$\alpha$ intensities of our model clouds are thus similar to those of observed C\hvcs. Note that we overestimate the intensities in our calculations, because we do not consider extinction in the line-of-sight. When comparing in detail to individual observed C\hvcs{} it is likely that thermal conduction alone is not a proper mechanism to reproduce real H$\alpha$ intensities. An additional mechanism to enhance H$\alpha$ intensity by heating up the clouds might be provided by a diffuse, soft X-ray background as being stated by \citet[][]{07bregman}. The cosmic UV background alone would account for only a negligible H$\alpha$ intensity of $5~$mR and is exceeded by the Galactic UV radiation field for distances from the disk of $\lesssim 100~$kpc \citep[][]{99maloneyblandhawthorn}.

\section{Summary and conclusions}\label{sec:sumcon}
We conducted 3D hydrodynamics simulations of multi-physical processes in high-velocity clouds by using the {\sc Flash} code. The capability of adaptive mesh refinement is extensively used with a finest numerical resolution of $1.3~$pc. By this, we are able to sufficiently resolve hydrodynamical instabilities (Kelvin-Helmholtz, Rayleigh-Taylor) as well as the Field length for thermal conduction. We tracked and compared the evolution of $14$ different model clouds which differ in mass, size, metallicity, and speed. All models are inititially in hydrostatic and thermal equilibrium internally and in pressure equilibrium with the ambient, hot medium. The physical processes of metal-dependent plasma cooling and heating, dissociation and ionization, and mass diffusion take place in all models. The model clouds either consider or exclude self-gravity and thermal conduction. Hence, we can directly compare the impact of cloud mass, speed, drag, metallicity, thermal conduction, and modeling simplicity on cloud evolution and discuss our findings in context of both their life-times and a potential gravitational collapse.

Ten out of the $14$ models survive the first $50~$Myr of their evolution (Table~\ref{tab:massloss-1}). These are all massive clouds with self-gravity (seven models) and three low-mass model clouds. To be not disrupted within $50~$Myr the low-mass clouds need thermal conduction, must not be too fast, and must not be too tiny in size (model L1\_me). Finally, four out of these $10$ models are not disrupted and in addition develop a local gravitational instability (Table~\ref{tab:grav-1}). These findings contradict to other studies, where clouds are only resistive to disruption for much higher masses \citep[][]{09heitschputman}, or larger radii and lower velocities \citep[][]{17armillottaetal}.

By being too simplistic in the modelling of clouds they are easily disrupted by Rayleigh-Taylor instabilities within $10$ sound-crossing times. Wrong conclusions may be drawn if self-gravity, thermal conduction, and a core-halo density profile are not accounted for.

The metallicity of clouds affects the strength of radiative cooling. A high metallicity of $\zsolar$ tends to produce cool clouds at mainly $10^2$ to $10^3~$K and with $0.1\zsolar$ the clouds are mainly at temperatures between $10^3$ to $10^4~$K. Since the observed core-halo structure in some C\hvcs{} reveals core temperatures of only af few $10^2~$K we conclude that metallicity is higher there, probably above $0.3\zsolar$.

It turns out that thermal conduction is a key process in the evolution of C\hvcs. Even though all of our simulated low-mass clouds will be disrupted, thermal conduction delays disruption by substantially lowering the mass-loss rate. Clumpy fragments of high density, which can cool efficiently, are stripped from clouds withouth thermal conduction while thermal conduction causes a continuous, low-density, filamentary stripping of gas at high temperatures, which is more likely to resemble the smooth density gradients observed between head and tail in some C\hvcs. We thus conclude that thermal conduction is at work in real C\hvcs.

Massive clouds are gravitationally stable for sufficiently low velocities (Mach~$\lesssim 0.5$), but for Mach~$\gtrsim 0.7$ they indeed form Jeans-instable regions. However, if thermal conduction is suppressed, the massive clouds can collapse even at Mach~$\sim 0.5$ (model M1\_tc) and are able to reach gas surface-densities ($0.16~$g~cm$^{-2}$) being high enough to account for a prospective star-formation site. We strongly suppose that thermal conduction is a key process in preventing real C\hvcs{} from forming stars.

It is not realistic, that clouds are not decelerated but keep constant speed, because the \cgm{} transfers momentum to the clouds. Combining both physical effects of life-prolonging thermal conduction and stripping-reducing drag (cloud M1\_dec) an instable region can form. The collapse commences not before the clouds are already decelerated to speeds representative for \ivcs. However, the decreasing ram-pressure is not sufficient to dominate the regions' thermal pressure maintained by thermal conduction. Hence, the instable region cannot be compacted to surface densities high enough for star formation.

\section*{Acknowledgements}
We gratefully acknowledge the very helpful comments by the anonymous referee that substantially improved the clarity of this paper. The authors gratefully acknowledge the valuable discussion by Nigel Mitchell, Sylvia Pl\"ockinger, and Wolfgang Vieser. This work was supported by the Doctoral College (Initiativkolleg) I033-N at the University of Vienna and by the Austrian Science Fund (FWF), project number P 21097. Furthermore, part of this work was realized during a research stay of BS at Jacobs University Bremen (Germany) by a grant from the KWA program of the University of Vienna (grant number 000386). BS wants to thank Elke R\"odiger and Marcus Br\"uggen for their hospitality and valuable input. The software used in this work was in part developed by the DOE NNSA-ASC OASCR Flash Center at the University of Chicago. The models have been computed at the Vienna Scientific Clusters 1 and 3 (VSC-1 and VSC-3)\footnote{see \href{http://vsc.ac.at/}{http://vsc.ac.at/}} and the Astro-Cluster at the Department of Astrophysics, University of Vienna.

\section*{Data availability}

The data underlying this article (simulation results, analysis scripts) will be shared on reasonable request to the corresponding author.

%%%%%%%%%%%%%%%%%%%%%%%%%%%%%%%%%%%%%%%%%%%%%%%%%%



\bibliographystyle{mnras}
\bibliography{references} % if your bibtex file is called example.bib

%%%%%%%%%%%%%%%%%%%%%%%%%%%%%%%%%%%%%%%%%%%%%%%%%%

%%%%%%%%%%%%%%%%% APPENDICES %%%%%%%%%%%%%%%%%%%%%

\appendix

\section{Dissociation and ionization in thermodynamic equilibrium}\label{app:dission}
At sufficiently high temperatures the electronic bond of H$_2$ can be broken with
\begin{equation}
y:=\frac{\varrho(\rm{H})}{\varrho(\rm{H})+\varrho(\rm{H}_2)}\label{equ:dission-1}
\end{equation}
being the fraction of dissociated H$_2$, where $\varrho$ denotes the respective mass density. In terms of the dissociation constant for H$_2$ \citep[][]{75blackbodenheimer} one obtains the conditional equation for $y$
\begin{equation}
\frac{y^2}{1-y}=\frac{2.11}{\varrho X}\exp\left\{\frac{-52,490}{T}\right\}=:F,\label{equ:dission-2}
\end{equation} 
where $X$ is the mass fraction of hydrogen. Solving Equ.~(\ref{equ:dission-2}) for $y$ yields the degree of dissociation
\begin{equation}
y(\varrho X,T)=-\frac F2+\sqrt{\frac{F^2}{4}+F}.\label{equ:dission-5}
\end{equation}
If temperatures increase above $T=E_{\rm ion}/k_B$, with $E_{\rm ion}=13.595~$eV being the ionization level, hydrogen is going to be \emph{ionized}. The necessary ionization energy can be provided by radiation (photons) or collisions (electrons). Vice versa, the ionized hydrogen can recombine again when it loses energy via radiation or collisions. Let $n_0$ and $n_1$ be the particle densities of neutral and ionized hydrogen, $n_{\rm e}$ be the electron number density, $R_{0\to 1}$ and $C_{0\to 1}$ be the rates of ionization by photons and collisions, and $R_{1\to 0}$ and $C_{1\to 0}$ be the respective recombination rates. We consider radiative ionization of hydrogen by soft X-rays with an ionization rate of $R_{0\to 1}^{\rm x}=7\times 10^{17}$ s$^{-1}$, and collisional ionization by both cosmic rays with an ionization rate of $C_{0\to 1}^{\rm cr}=7\times 10^{17}$ s$^{-1}$ and electrons with an ionization rate of $C_{0\to 1}^{\rm el}=n_{\rm e}Q_{0\to 1}^{\rm el}$. The rate coefficient for collisional ionization by electrons
\begin{eqnarray}
Q_{0\to 1}^{\rm el}(T)&=&5.465\times 10^{-11}\sqrt{T}\times\exp\left\{-\frac{13.543\;{\rm eV}}{kT}\right\}\times \nonumber\\
&&(-0.435+0.3\log T)\;{\rm cm}^3\;{\rm s}^{-1}.\label{equ:dission-7}
\end{eqnarray}
is taken from \citet{67mihalas}.

On the other hand, we consider recombination by radiation with a rate $R_{1\to 0}^{\rm rad}=n_{\rm e}\sum_{i=2}^\infty\alpha_i$. Captures to level $i=1$ are excluded, since the released photons do not contribute to the net recombination (``on the spot approximation''). Hence, only recombination coefficient \citep[e.g.][]{98spitzer}
\begin{equation}
\alpha_2(T)=\frac{2.06\times 10^{-11}Z^2}{\sqrt{T}}\phi_2\left(\frac{h\nu}{kT}\right)\;{\rm cm}^3\;{\rm s}^{-1},\label{equ:dission-8}
\end{equation}
which includes all recombinations to level $i=2$, is taken into account. The atomic number $Z=1$ for hydrogen. The function $\phi_2$ slowly varies with $T$ and is tabulated in \citet{98spitzer}. So, in thermodynamic equilibrium,
\begin{equation}
n_0\left(R_{0\to 1}^{\rm x}+C_{0\to 1}^{\rm cr}+n_{\rm e}Q_{0\to 1}^{\rm el}\right)=n_1n_{\rm e}\alpha_2,\label{equ:dission-9}
\end{equation}
with $n_{\rm e}$ being approximated by $n_{\rm e}\approx n_1+\chi_0(n_0+n_1)$, where the second summand counts free electrons being released by species having lower ionization levels than hydrogen. The fraction of these species is given by $\chi_0=5\times 10^{-4}$ \citep{72dalgarnomccray}. Let
\begin{equation}
f:=\frac{n_1}{n_0+n_1}\label{equ:dission-10}
\end{equation}
be the fraction of ionized hydrogen. By substituting Equ.~(\ref{equ:dission-10}) into Equ.~(\ref{equ:dission-9}) one obtains
\begin{equation}
0=f^2+Af-B,\label{equ:dission-11}
\end{equation}
with
\begin{eqnarray}
A&:=&\left[\chi_0\alpha_2+(\chi_0-1)Q_{0\to 1}^{\rm el}+\frac{R_{0\to 1}^{\rm x}+C_{0\to 1}^{\rm cr}}{n_0}\right]\left[\frac{1}{\alpha_2+Q_{0\to 1}^{\rm el}}\right], \nonumber \\
B&:=&\left[\frac{R_{0\to 1}^{\rm x}+C_{0\to 1}^{\rm cr}}{n_0}+\chi_0Q_{0\to 1}^{\rm el}\right]\left[\frac{1}{\alpha_2+Q_{0\to 1}^{\rm el}}\right]
\end{eqnarray}
are subsuming the rates of ionization and recombination. Solving Equ.~(\ref{equ:dission-11}) for $f$ yields the degree of ionization 
\begin{equation}
f(n_0,T)=-\frac A2+\sqrt{\frac{A^2}{4}+B}.\label{equ:dission-13}
\end{equation}

\section{Code test for implementation of saturated thermal conduction}\label{app:codetest}
\begin{figure}
\centering
\includegraphics[width=0.5\textwidth]{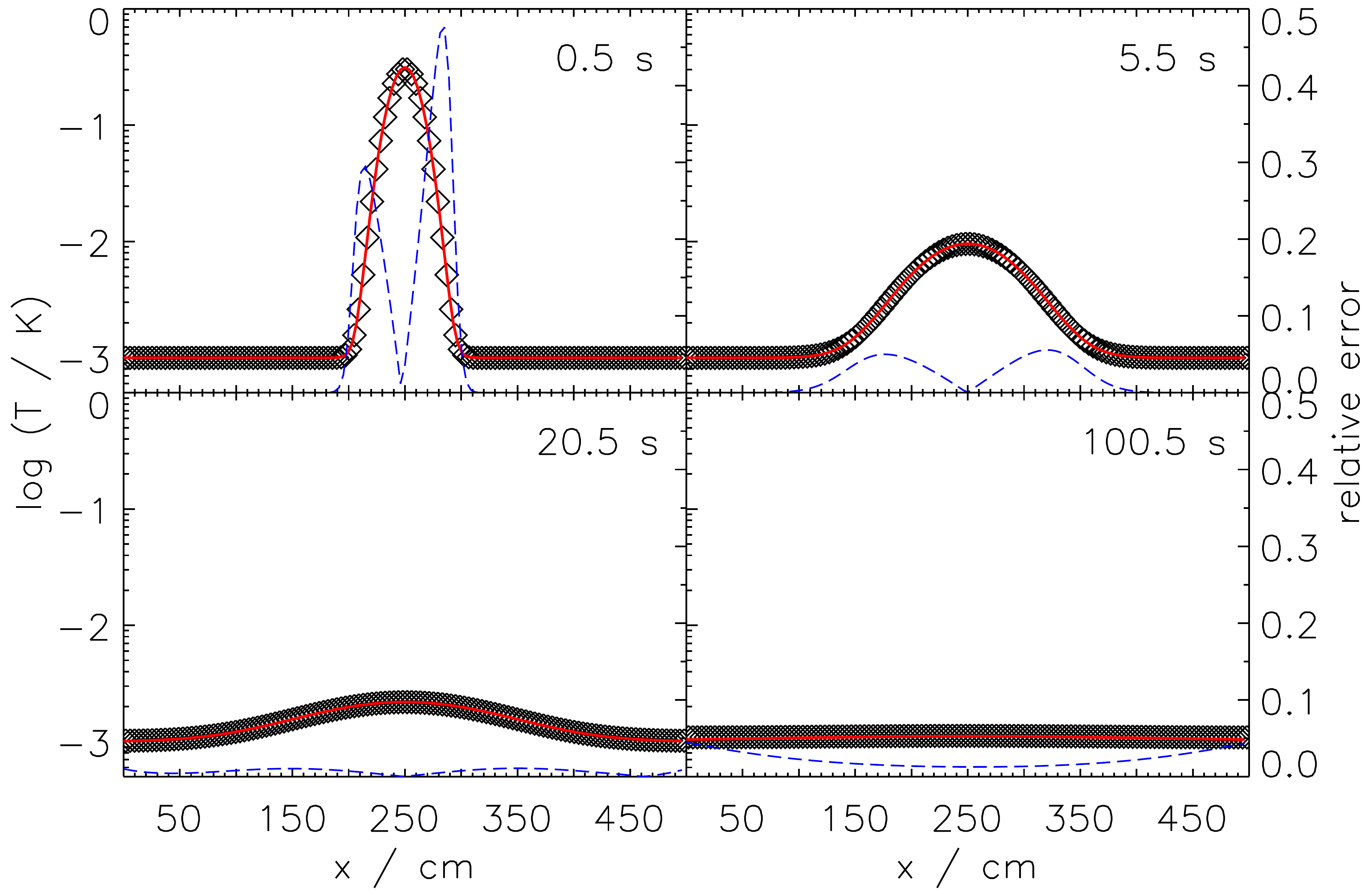}
\caption{Numerically obtained evolution of a three-dimensional Gaussian temperature distribution being subject to diffusion only (\emph{black diamonds}) compared with the analytical solution (\emph{red solid line}) at different times. The \emph{blue dashed line} shows the relative error between both solutions. The temperature profile is shown along the x-direction.}
\label{fig:codetest-1}
\end{figure}

We are going to show the evolution of a Gaussian temperature distribution being subject to diffusion only\footnote{This test problem comes along with the {\sc Flash} code and is adapted accordingly by us.}. For any twice differentiable, scalar function $f(x,y,z,t)$, $x,y,z,t\in \mathbb{R}$, $f:\mathbb{R}^3\times\mathbb{R}\to\mathbb{R}^3\times\mathbb{R}$, satisfying the heat equation
\begin{equation}
\frac{\partial f}{\partial t}-\left(\alpha_x\frac{\partial^2f}{\partial x^2}+\alpha_y\frac{\partial^2f}{\partial y^2}+\alpha_z\frac{\partial^2f}{\partial z^2}\right)=0,\label{equ:app-codetest-1}
\end{equation}
a delta distribution at any initial time $t_0$
\begin{equation}
f(x,y,z,t_0)=Q(t_0)\delta(x,y,z)\label{equ:app-codetest-2}
\end{equation}
spreads into a Gaussian distribution
\begin{eqnarray}
f(x,y,z,t)&=&\frac{Q(t_0)}{[4\pi(t-t_0)]^{3/2}(\alpha_x\alpha_y\alpha_z)^{1/2}} \nonumber\\
&\times&\exp\left\{-\frac{1}{4(t-t_0)}\left(\frac{x^2}{\alpha_x}+\frac{y^2}{\alpha_y}+\frac{z^2}{\alpha_z}\right)\right\}\label{equ:app-codetest-3}
\end{eqnarray}
and remains a Gaussian distribution for all times $t>t_0$. The analytical result~(\ref{equ:app-codetest-3}) is well-known. By identifying $f(x,y,z,t)$ with the temperature field of a fluid and taking $\alpha=\alpha_x=\alpha_y=\alpha_z$ (isotropic diffusion), we obtain the evolution as being determined by Equ.~(\ref{equ:app-codetest-3}). We point out that $\alpha$ is a function of the effective conductivity $\kappa_{\rm eff}$~(\ref{equ:satcond-5}). We have chosen $Q(t_0)=10^4~$K~cm$^3$ at $t_0=0.5~$s (to avoid numerical difficulties implied by the Delta function at $t=0$) and a both spatially and temporally constant density of $\varrho=1$~g~cm$^{-3}$. The centre of the distribution is located at $(x_{\rm c},y_{\rm c},z_{\rm c})=(250,250,250)~$cm and the side length of the computational domain is $500~$cm. The domain is discretized by $128$ cells per dimension yielding a grid spacing of $\Delta x=3.9~$cm. We use periodic boundary conditions. The background temperature is fixed at $10^{-3}~$K. We find $f(x_{\rm c},y_{\rm c},z_{\rm c},t_0)=0.3~$K accounting for an initial temperature contrast of $\Delta T=300$. In Fig.~\ref{fig:codetest-1} the analytical solution~(\ref{equ:app-codetest-3}) is compared to the numerical one obtained by the {\sc Flash} code. Initially, the relative error is highest where the Gaussian temperature distribution passes into the background temperature: the change in slope cannot be approximated sufficiently by the grid, i.e. the numerical resolution is to coarse. However, the relative error is lowered to a reasonable level after $10$ diffusion times ($\tau=0.5~$s). 
Boundary effects, which may superpose the numerical solution, are not observed. % Hence, a sufficient accuracy of the numerical solution is deduced.
The numerical solution is hence sufficiently accurate.
%
% Don't change these lines
\bsp	% typesetting comment
\label{lastpage}
\end{document}